\documentclass[manuscript,screen, nonacm]{acmart}

\AtBeginDocument{%
  \providecommand\BibTeX{{%
    \normalfont B\kern-0.5em{\scshape i\kern-0.25em b}\kern-0.8em\TeX}}}

\setcopyright{acmcopyright}
\copyrightyear{2024}
\acmYear{2024}
\acmDOI{XXXXXXX.XXXXXXX}

\acmJournal{CSUR}

\usepackage{listings}

\usepackage{xcolor}
\usepackage{mathtools}
\usepackage{subcaption}
\usepackage{multirow}
\usepackage{caption}

\usepackage[ruled, lined, linesnumbered, commentsnumbered, longend]{algorithm2e}

\SetCommentSty{mycommfont}

\definecolor{codegreen}{rgb}{0,0.6,0}
\definecolor{codegray}{rgb}{0.5,0.5,0.5}
\definecolor{codepurple}{rgb}{0.58,0,0.82}
\definecolor{backcolour}{rgb}{0.95,0.95,0.95}

\definecolor{delim}{RGB}{20,105,176}
\definecolor{numb}{RGB}{106, 109, 32}
\definecolor{string}{rgb}{0.64,0.08,0.08}

\lstdefinestyle{qfaasstyle}{
    backgroundcolor=\color{backcolour},   
    commentstyle=\color{codegreen},
    keywordstyle=\color{magenta},
    numberstyle=\tiny\color{codegray},
    stringstyle=\color{codepurple},
    basicstyle=\ttfamily\footnotesize,
    breakatwhitespace=false,         
    breaklines=true,                 
    captionpos=b,                    
    keepspaces=true,                 
    numbers=left,                    
    numbersep=10pt,                  
    showspaces=false,                
    showstringspaces=false,
    showtabs=false,                  
    tabsize=2,
    frame=single,
    xleftmargin=0.5in,
    xrightmargin=.25in
}

\lstdefinelanguage{json}{
    rulecolor=\color{black},
    postbreak=\raisebox{0ex}[0ex][0ex]{\ensuremath{\color{gray}\hookrightarrow\space}},
    upquote=true,
    morestring=[b]",
    literate=
     *{0}{{{\color{numb}0}}}{1}
      {1}{{{\color{numb}1}}}{1}
      {2}{{{\color{numb}2}}}{1}
      {3}{{{\color{numb}3}}}{1}
      {4}{{{\color{numb}4}}}{1}
      {5}{{{\color{numb}5}}}{1}
      {6}{{{\color{numb}6}}}{1}
      {7}{{{\color{numb}7}}}{1}
      {8}{{{\color{numb}8}}}{1}
      {9}{{{\color{numb}9}}}{1}
      {\{}{{{\color{delim}{\{}}}}{1}
      {\}}{{{\color{delim}{\}}}}}{1}
      {[}{{{\color{delim}{[}}}}{1}
      {]}{{{\color{delim}{]}}}}{1},
}

\begin{document}

\title{Quantum Cloud Computing: A Review, Open Problems, and Future Directions}

\author{Hoa T. Nguyen}
\email{thanhhoan@student.unimelb.edu.au}
\orcid{0000-0001-6904-6312}
\affiliation{%
  \institution{Cloud Computing and Distributed Systems (CLOUDS) Laboratory, School of Computing and Information Systems, The University of Melbourne}
  \city{Parkville}
  \state{Victoria}
  \country{Australia}}

\author{Prabhakar Krishnan}
\email{kprabhakar@am.amrita.edu}
\orcid{0000-0001-6702-112X}
\affiliation{%
  \institution{Center for Cybersecurity Systems and Networks, Amrita Vishwa Vidyapeetham}
  \city{Amritapuri Campus}
  \state{Kerala}
  \country{India}}

\author{Dilip Krishnaswamy}
\email{dilip@ieee.org}
\orcid{0000-0002-3808-7733}
\affiliation{%
  \institution{Quantum Walks Technologies (QWalks)}
  \city{New Jersey}
  \state{NJ}
  \country{USA}}

\author{Muhammad Usman}
\orcid{0000-0003-3476-2348}
\affiliation{%
  \institution{School of Physics, The University of Melbourne}
  \city{Parkville}
  \state{Victoria}
  \country{Australia}}
  \affiliation{%
  \institution{Data61, CSIRO}
  \city{Clayton}
  \state{Victoria}
  \country{Australia}}
\email{musman@unimelb.edu.au}

\author{Rajkumar Buyya}
\orcid{0000-0001-9754-6496}
\affiliation{%
  \institution{Cloud Computing and Distributed Systems (CLOUDS) Laboratory, School of Computing and Information Systems, The University of Melbourne}
  \city{Parkville}
  \state{Victoria}
  \country{Australia}}
\email{rbuyya@unimelb.edu.au}

\renewcommand{\shortauthors}{H. Nguyen, P. Krishnan, D. Krishnaswamy, M. Usman, and R. Buyya}

\begin{abstract}

Quantum cloud computing is an emerging paradigm of computing that empowers quantum applications and their deployment on quantum computing resources without the need for a specialized environment to host and operate physical quantum computers. This paper reviews recent advances, identifies open problems, and proposes future directions in quantum cloud computing. It discusses the state-of-the-art quantum cloud advances, including the various cloud-based models, platforms, and recently developed technologies and software use cases. 
Furthermore, it discusses different aspects of the quantum cloud, including resource management, quantum serverless, security, and privacy problems. Finally, the paper examines open problems and proposes the future directions of quantum cloud computing, including potential opportunities and ongoing research in this emerging field. 

\end{abstract}



\maketitle

\section{Introduction}
Quantum computing is anticipated to revolutionize many scientific fields by promising to solve intractable computational problems beyond the capabilities of the state-of-the-art classical computers. Although quantum computers are still in the early stages of development, they have already been used to address critical problems, such as simulating the behavior of molecules \cite{gobato_calculations_2017, magann_digital_2021} and discovering new drugs \cite{cao_potential_2018,zinner_quantum_2021,batra_quantum_2021}. However, acquiring a dedicated physical quantum computer and hosting it is challenging as they require an extremely special environment to host and operate, such as extremely cold temperatures and complicated controlling \cite{saki_survey_2021}. Due to the complexity of operating a physical quantum computer, hosting them within cloud computing environments is the most common, as they provide easy access to today's quantum computation resources. Thus, quantum machines are placed in a remote data center with particular environmental conditions. These devices can interact with end-users via the vendor cloud platform, which allows for the execution of quantum circuits and the retrieval of the computation result. Engineers and researchers, therefore, have more convenient ways to access and utilize quantum devices remotely from their local computers. This model shortens the pathway to achieving quantum advantages in the current era of noisy intermediate-scale quantum (NISQ) hardware. NISQ devices, characterized by their tens to hundreds of qubit scale, operate with a significant level of noise and limited qubit coherence times, which presents unique challenges for quantum computation \cite{nisq-preskill}. Despite these limitations, NISQ technology offers a promising platform for exploring quantum algorithms and applications prior to the advent of fault-tolerant quantum computing.

The emerging field of quantum cloud computing (QCC) combines the principles of quantum computing with cloud infrastructure, enabling remote access to quantum computers. This integration promises to significantly lower the barrier to utilizing quantum computing resources, making it feasible for researchers and developers to explore quantum algorithms without the need for their own quantum hardware. Given the rapid advancements in QCC, which democratize access to quantum computational power and facilitate a wide range of applications from cryptography \cite{huang_quantum_2021, li_quantum_2021, kumar_quantum_2022}, quantum machine learning \cite{berganza_gomez_towards_2022, gong_quantum_2021, Fadli2023-ad, hibatallah2023framework}, to complex system simulations \cite{gobato_calculations_2017, magann_digital_2021}, there is a pressing need for a comprehensive review of the current research and developments in this area. This survey aims to examine the progress of QCC through the perspective of corresponding concepts and problems in classical cloud computing and provide a useful resource for researchers in both quantum computing and cloud computing domains. 
This paper organizes the key aspects of QCC into five areas: 1) computing models, 2) software use cases, 3) providers and platforms, 4) resource management, and 5) security and privacy. For each aspect, we present a summary of the pioneering work, its relationship to classical models, its variants, and its potential applications in advancing practical research. To the best of our knowledge, this is one of the first reviews of QCC, and we hope it will serve as a valuable guide for researchers and practitioners in the domain.

\subsection{Contributions}
As one of the first comprehensive review on quantum cloud computing, the major contributions and novelty of our review article are:
\begin{itemize}
    \item We introduce the concepts and survey recent studies of quantum cloud computing and its emerging models, including hybrid quantum-classical computing and quantum serverless. We summarize key industrial players in quantum cloud computing and recent developments in the field.
    
    \item We provide a comprehensive overview of the state-of-the-art, including the various technologies, platforms, and applications that have been developed for quantum cloud computing. Our survey presents various applications of quantum cloud computing in different fields, such as machine learning and cryptography. We also review quantum cloud providers and software platforms for utilizing quantum cloud resources.

    \item We review recent advances in quantum cloud computing in emerging aspects, including quantum cloud resource management, distributed quantum computation techniques, and quantum cloud security.

    \item We identify key challenges and open problems that need to be overcome, ongoing research, and propose future research directions to realize the potential of quantum cloud computing fully.
\end{itemize}

\subsection{Related surveys}
In the literature, numerous review papers and surveys focus on quantum computation and information, as summarized in Table \ref{tab:related-work}. Some surveys also cover the communication and networking aspects of quantum computing. However, very few papers are solely dedicated to quantum cloud computing. Moreover, the reviews that include this topic often provide only a superficial discussion or lack a detailed analysis of its recent advances and open problems. To bridge the gap, our work focuses on reviewing different aspects of recent quantum cloud computing development, including its architectures, computation models, applications, security, and relevant open research problems.

\begin{table}[htbp]
\caption{Summary of related surveys on quantum cloud computing 
(QC: Quantum computing, QS: Quantum software, QML: Quantum machine learning, QN: Quantum Network, QDS: Quantum Distributed Systems)}
\label{tab:related-work}
\begin{tabular}{|l|l|l|l|}
\hline
\textbf{Survey} & \textbf{Focus} & \textbf{Topic covered} & \textbf{Year} \\ \hline
Gyongyosi et al. \cite{gyongyosi_survey_2019} & \multirow{4}{*}{QC} & QC fundamentals and quantum algorithm implementations & 2019 \\ \cline{1-1} \cline{3-4} 
Gill et al. \cite{gill_quantum_2021} &  & Taxonomy, systematic review and future directions of QC & 2021 \\ \cline{1-1} \cline{3-4} 
Resch and Karpuzcu \cite{resch_survey_benchmark_2021} &  & QC benchmarking, quantum noise & 2021 \\
\cline{1-1} \cline{3-4} 
Yang et al. \cite{yang_survey_2023} &  & Quantum hardware, quantum networks, quantum cryptography, QML & 2023
\\ \hline
Serrano et al. \cite{serrano_quantum_2022} & \multirow{3}{*}{QS} & QS components, platforms, and quality assessments & 2022 \\ \cline{1-1} \cline{3-4}
Moguel et al. \cite{moguel_quantum_2022} &  & Service-oriented QS, practical implementation, and limitations & 2022 \\ \cline{1-1} \cline{3-4}
Khan et al. \cite{KHAN2023111682} &  & QS architectures, tools, and frameworks & 2023 \\ \hline
O'Quinn and Mao \cite{oquinn_quantum_2020} & \multirow{4}{*}{QML} & QML algorithms, quantum cloud platforms, quantum annealing & 2020 \\ \cline{1-1} \cline{3-4} 
Zhang et al. \cite{zhang_recent_2020} &  & QML algorithms and applications & 2020 \\ \cline{1-1} \cline{3-4}
Massoli et al. \cite{massoli_survey_2022} &  & QC, quantum perceptrons, quantum neural networks  & 2022 \\ \cline{1-1} \cline{3-4}
Tian et al. \cite{tian_recent_2022} &  & Quantum Generative Machine Learning models & 2023 \\ \hline
Caleffi et al. \cite{caleffi_quantum_2018} & \multirow{5}{*}{QN} & Quantum Internet & 2018 \\ \cline{1-1} \cline{3-4} 
Wehner et al. \cite{wehner_quantum_2018} &  & Quantum Internet & 2018 \\ \cline{1-1} \cline{3-4} 
Cacciapuoti et al. \cite{cacciapuoti_quantum_2020} &  & Quantum Internet & 2020 \\ \cline{1-1} \cline{3-4} 
Mehic et al. \cite{mehic_quantum_2020} &  & Network aspects for Quantum Key Distribution & 2020 \\ \cline{1-1} \cline{3-4} 
Fang et al. \cite{fang_quantum_2023} &  & QN components, devices, protocols, applications and toolkits & 2023 
\\ \hline
Kaiiali et al. \cite{kaiiali_cloud_2019} &  \multirow{5}{*}{QDS} & Impact of QC on cloud computing and vice versa & 2019 \\ \cline{1-1} \cline{3-4} 
Cuomo et al. \cite{cuomo_towards_2020} &  & Vision for quantum distributed system & 2020 \\ \cline{1-1} \cline{3-4} 
Leymann et al. \cite{leymann_quantum_2020} &  & Applications, use cases and collaborative quantum platforms & 2020
\\ \cline{1-1} \cline{3-4} 
Loke \cite{loke_distributed_2022} &  & QDS, quantum network, quantum Internet computing & 2022 \\ \cline{1-1} \cline{3-4} 
Caleffi et al. \cite{caleffi_distributed_2022} &  & Networking, compiling, and simulation approaches for QDS & 2022 
\\ \cline{1-1} \cline{3-4} 
\textbf{\textit{Our work - This paper}} &  & \begin{tabular}[c]{@{}l@{}}Concepts, architecture models, software and frameworks, resource \\management, security and privacy of quantum cloud computing\end{tabular}  & 2024 

 \\ \hline
\end{tabular}
\end{table}

There has been tremendous work in recent years regarding general quantum computing surveys. Gyongyosi et al. \cite{gyongyosi_survey_2019} reviewed the fundamentals of quantum computing, large-scale computing models, and recent quantum algorithm implementations. Similarly, Gill et al. \cite{gill_quantum_2021} proposed a taxonomy of quantum computing, a systematic review of recent progress, and the future direction of quantum computing in general. Although they briefly mention quantum cloud computing as a potential direction, no detailed analysis or discussion was provided. From a practical perspective, Resch and Karpuzcu \cite{resch_survey_benchmark_2021} reviewed different benchmarking approaches for quantum systems and investigated the impact of quantum noise on the performance of quantum computers. Subsequently, Yang et al. \cite{yang_survey_2023} provided a comprehensive survey to illustrate the big picture of key challenges in quantum computing and communications, focusing on quantum hardware, quantum networks, quantum cryptography, and quantum machine learning. 

In terms of quantum software engineering, Serrano et al. \cite{serrano_quantum_2022} reviewed the quantum software components and platforms and proposed the quality assessment requirements for quantum software. Similarly, Moguel et al. \cite{moguel_quantum_2022} reviewed the current state of quantum software engineering, focusing on service-oriented quantum computing, with several case study implementations on Amazon Braket. Additionally, Khan et al. \cite{KHAN2023111682}
provided a comprehensive study of the architectures, frameworks, and tools for quantum software engineering.
In addition to the core topics of quantum computing, Quantum Machine Learning (QML) has emerged as a pivotal area of research, attracting substantial attention for its potential to revolutionize computational paradigms. Several surveys have systematically cataloged the advancements and applications of QML algorithms. O'Quinn and Mao \cite{oquinn_quantum_2020} provided a brief overview of quantum machine learning algorithms and existing quantum cloud platforms for quantum applications. Zhang et al. \cite{zhang_recent_2020} provided a comprehensive review of the state-of-the-art in quantum machine learning algorithms, highlighting their applications and the theoretical underpinnings that facilitate these advancements. Massoli et al. \cite{massoli_survey_2022} review recent progress in theoretical formulations, simulations, and implementations of quantum perceptrons and quantum neural networks.
Similarly, Tian et al. \cite{tian_recent_2022} offered a focused review on the advancement within quantum generative learning models, outlining the recent developments and their implications for the broader field of quantum computing.

From a distributed computing perspective, many studies reviewed quantum networks and the quantum Internet, focusing on quantum communication, security, and networking aspects. 
Caleffi et al. \cite{caleffi_quantum_2018} briefly discussed several research problems for quantum Internet design and development. 
Similarly, Wehner et al. \cite{wehner_quantum_2018} provided an overview of the current stage and outlook on the quantum Internet, highlighting the requirement to build a reliable large-scale quantum network in the future. 
Subsequently, Cacciapuoti et al. \cite{cacciapuoti_quantum_2020} review the quantum Internet's key principles and network challenges.
Mehic et al. \cite{mehic_quantum_2020} reviewed the recent development of the quantum key distribution, focusing on networking aspects.
Similar surveys by Singh et al. \cite{singh_quantum_internet_2021} and Krishnamurthi et al. \cite{kumar_comprehensive_2022} also comprehensively review the recent advancement of quantum Internet regarding its applications, functionalities, and technologies. Illiano et al. \cite{illiano_quantum_2022} provided the state-of-the-art quantum Internet protocol stack and outlined several research directions.
Some review and vision articles also focused on distributed quantum computing, such as \cite{cuomo_towards_2020, loke_distributed_2022, caleffi_distributed_2022}.
Cuomo et al. \cite{cuomo_towards_2020} discussed the vision for a future distributed quantum computing ecosystem, with quantum Internet as the key underlying infrastructure.
Loke \cite{loke_distributed_2022} devised an overview of quantum distributed computing and quantum networking toward quantum Internet computing in the future. In \cite{caleffi_distributed_2022}, Caleffi et al. provided an overview of recent advances and key challenges of distributed quantum computing concerning quantum networking, quantum compiling, and simulation approaches. 
Fang et al. \cite{fang_quantum_2023} provided a holistic review of different aspects of quantum networks from both theoretical and practical perspectives. This study discusses the recent development of quantum network components, physical devices, protocols, applications, and network toolkits for scientific research. A few existing studies paid attention to quantum computing's impact on cloud computing. Kaiiali et al. \cite{kaiiali_cloud_2019} reviewed the potential impact of quantum computing on cloud computing and vice versa. However, this article focuses on the security aspect and some high-level impact of these two areas on each other without reviewing the recent advances, opportunities, and open challenges of quantum cloud computing as our survey. Leymann et al. \cite{leymann_quantum_2020} reviewed some application potentials and opportunities of quantum computing on the cloud, focusing on software tools and platforms for collaborating to develop quantum applications. In \cite{soeparno_cloud_2021}, Soeparnoa et al. conducted some experiences on IBM Quantum and Qutech cloud services and gave some general overview and comparison between these quantum cloud platforms. 

However, as far as we are aware, none of the existing surveys focused explicitly on recent advances in quantum cloud computing, such as the quantum cloud service model, quantum serverless, resource management, cloud-based platforms, and related open problems.
Therefore, our paper can be considered one of the first surveys focusing on different aspects of quantum cloud computing. It provides a thorough review of recent advances, research challenges, and open problems in this emerging yet crucial research area.

\begin{figure}[htbp]
    \centering
    \includegraphics[scale=0.14]{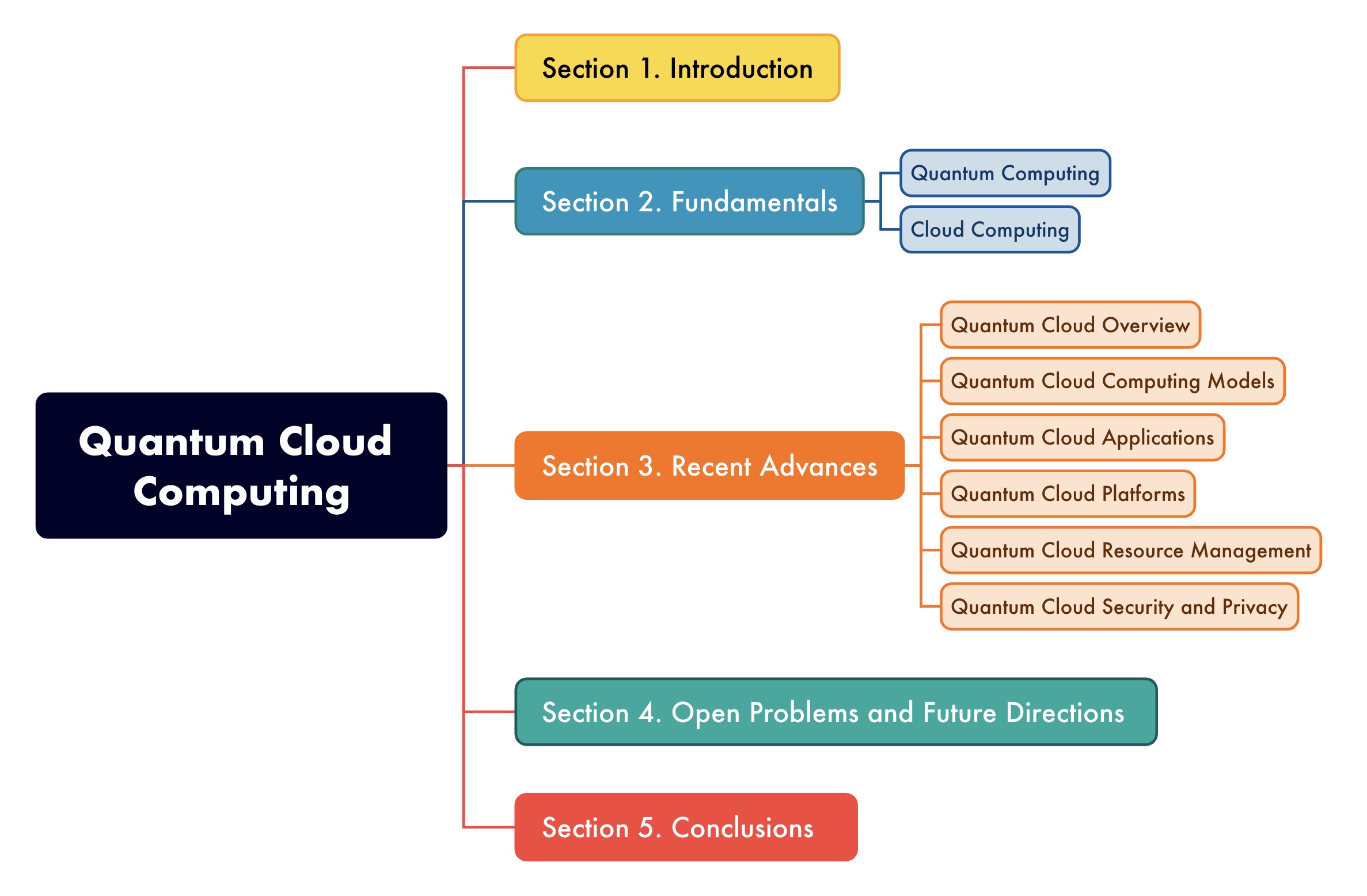}
    \caption{The graphical outline of the survey}
    \label{fig:survey-outline}
\end{figure}

\subsection{Structure of the survey}
The structure of this survey is illustrated in Figure \ref{fig:survey-outline}, and the rest of the paper is organized as follows: Section 2 covers the foundational concepts of quantum computing and cloud computing to accommodate the diversity of readers' backgrounds. In Section 3, we comprehensively review recent research on quantum cloud computing and examine their connections to classical cloud computing models. This section is divided into six sub-sections, focusing on quantum cloud computing concepts and providers, quantum cloud applications, quantum serverless, resource management problems, distributed quantum computing, and quantum cloud security. Then, we discuss the challenges and potential future directions in the field of quantum cloud computing in section 4. Finally, we conclude and summarize our survey in section 5.

\section{A Synopsis of Quantum Computing and Cloud Computing}
This section provides a brief overview of quantum computing and cloud computing to take into account the diverse backgrounds of readers.

\subsection{Quantum Computing}
Quantum computing is an emerging paradigm that utilizes quantum-mechanical phenomena, such as superposition and entanglement, to perform operations on data. It has the potential to solve certain problems such as optimization \cite{qalgo-qaoa, lewis2017quadratic, kochenberger2014unconstrained}, financial modeling \cite{herman2023quantum, orus2019quantum}, molecule simulation \cite{gobato_calculations_2017, magann_digital_2021}, and machine learning \cite{west2023towards, berganza_gomez_towards_2022, gong_quantum_2021, Fadli2023-ad, hibatallah2023framework} much faster than classical computers. The subsequent discussion focuses on the cornerstone of quantum computing: the quantum bit, computation models, and quantum algorithms.

\subsubsection{Quantum bits (Qubits)} A quantum bit (or qubit) is the basic unit of quantum information. Different from classical bits, which can exist in one of two states (0 or 1), qubits are characterized by their ability to exist in a superposition state, i.e., they can be in a combination of both states 0 and 1 simultaneously. This property enables the representation and processing of a richer set of information possibilities compared to classical bits. Another key characteristic of qubits is entanglement, a quantum phenomenon where the state of one qubit becomes inextricably linked to the state of another, regardless of the distance between them. This linkage means the properties of entangled qubits cannot be described independently of each other \cite{qbook-nielsen-chuang}. These distinctive features of qubits could enable quantum computers to execute certain computations more efficiently than classical computers, potentially solving problems considered intractable for classical computation. 


\begin{table*}[htbp]
  \small
  \caption{A summary of common quantum gates and their mathematical representation}
  \label{tab:qgates}
  \renewcommand{\arraystretch}{1.3}
  \centering
  \begin{tabular}{|p{0.02\textwidth}|p{0.13\textwidth}|p{0.25\textwidth}|p{0.51\textwidth}|}
    \hline
    \multicolumn{2}{|l|}{\textbf{Quantum gates}}  & \textbf{Description}  &  \textbf{Mathematical Representation} \\
    \hline
    \multirow{8}{*}{\rotatebox[origin=c]{90}{\textbf{Single-qubit}}} & Pauli-X & A bit flip gate, which acts as a rotation by $\pi$ around the X-axis of the Bloch sphere & $\sigma_X =  X = \begin{bmatrix} 0 & 1 \\ 1 & 0 \end{bmatrix} = |0\rangle\langle1| + |1\rangle\langle0|$    \\
    & Pauli-Y & A bit and phase flip gate, which acts as a rotation by $\pi$ around the Y-axis of the Bloch sphere & $\sigma_Y = Y = \begin{bmatrix} 0 & -i \\ i & 0 \end{bmatrix} = -i|0\rangle\langle1| + i|1\rangle\langle0| = i.\sigma_X.\sigma_Z$  \\
    & Pauli-Z & A phase flip gate, which acts as a rotation by $\pi$ around the Z-axis of the Bloch sphere  & $\sigma_Z = Z = \begin{bmatrix} 1 & 0 \\ 0 & -1 \end{bmatrix} = |0\rangle\langle0| - |1\rangle\langle1|$   \\
    & Hadamard (H) & A gate to create a superposition of $|0\rangle$ and $|1\rangle$ & $H = \tfrac{1}{\sqrt{2}}\begin{bmatrix} 1 & 1 \\ 1 & -1 \end{bmatrix} = \tfrac{1}{\sqrt{2}}(|0\rangle\langle0|+|0\rangle\langle1| + |1\rangle\langle0| - |1\rangle\langle1|)$   \\
    & Phase ($P_\phi$) & A parametrized gate performs a rotation of $\phi$ around the Z-axis direction & $P(\phi) = \begin{bmatrix} 1 & 0 \\ 0 & e^{i\phi} \end{bmatrix}$ where $\phi$ is a real number   \\
    & Identity (I) & A gate that have no effect on the qubit state & $I = \begin{bmatrix} 1 & 0 \\ 0 & 1\end{bmatrix}$   \\
    & S (or $\sqrt{Z} $) & A $P(\phi)$ gate with $\phi = \pi/2$, apply a quarter-turn around the Bloch sphere& $S = \begin{bmatrix} 1 & 0 \\ 0 & i\end{bmatrix} = \begin{bmatrix} 1 & 0 \\ 0 & e^{\frac{i\pi}{2}} \end{bmatrix}$   \\
    & T (or $\sqrt[4]{Z}$) & A $P(\phi)$ gate with $\phi = \pi/4$ & $T = \begin{bmatrix} 1 & 0 \\ 0 & e^{\frac{i\pi}{4}} \end{bmatrix}$   \\
    \hline
    \multirow{3}{*}{\rotatebox[origin=c]{90}{\textbf{Multi-qubit \quad}}} 
    & CNOT (or CX) & A 2-qubit gate that flips the target qubit when the control qubit is in the state $|1\rangle$  & $CX = \begin{bmatrix}1 &0 &0 &0 \\0&1&0&0\\0&0&0&1\\0&0&1&0\end{bmatrix} = |00\rangle\langle00| + |01\rangle\langle01|+|10\rangle\langle11|+|11\rangle\langle10|$   \\
    & SWAP & A 2-qubit gate that swap (exchange) the state of these two qubits & $SWAP =
    \begin{bmatrix}
        1 & 0 & 0 & 0 \\
        0 & 0 & 1 & 0 \\
        0 & 1 & 0 & 0 \\
        0 & 0 & 0 & 1
    \end{bmatrix}$    \\
    & Toffoli (or CCX) & A 3-qubit gate that flips the third qubit if the first two qubits are both in state $|1\rangle$ & $CCX q_0, q_1, q_2 =
    I \otimes I \otimes |0 \rangle \langle 0| + CX \otimes |1 \rangle \langle 1|$   \\
  \hline
\end{tabular}
\end{table*}

\subsubsection{Quantum computation models}
Two distinct quantum computing models are being concurrently developed: gate-based computations and quantum annealing-based computations. 
Gate-based models, also known as circuit-based models, use a set of quantum gates to perform operations on qubits. These gates are used to manipulate the state of qubits and perform operations on them. 
Table \ref{tab:qgates} briefly summarizes typical quantum gates' characteristics. Gate-based models are based on building a quantum circuit to realize a sequence of gates that perform a specific computation. The action of quantum gates can be represented as a unitary transform on qubits. If a sequence of gates $G_1$, $G_2$, \ldots $G_m$ are applied to a set of input qubit state vector $Q$, then the resultant state is given by the corresponding cascaded sequence of unitary transforms applied to the input vector $Q$ given by the matrix product $ G_m \ldots G_2 G_1 Q$. This approach allows for flexibility and can be used to support a wide range of quantum algorithms, such as search and factorization problems.

Quantum annealing-based computations support an alternate paradigm of computation, and they are typically utilized to solve optimization problems. In this approach, an energy function (also known as the Hamiltonian of the system) is used to represent an optimization problem under consideration. For example, a QUBO (Quadratic Unconstrained Binary Optimization) \cite{lewis2017quadratic} formulation involves self-coupling weights $w_{ii}$ for qubit states $x_i$ and cross-coupling weight terms $w_{ij}$ for pairwise products of qubit state terms  $x_i x_j$ to result in an energy function of the form $H = \sum_i w_{ii} x_i  + \sum_{ij} w_{ij} x_i x_j$. Many optimization problems can be reduced to such a form for the energy function that is desired to be minimized. There are physical self-coupling and cross-coupling connections in an annealing-based quantum computer. In some simple cases, the optimization problem can be directly mapped to the physical networked connections in the quantum computer. In other cases where this is not feasible (for example, where the desired degree of cross-coupled connectivity is higher than the physically available connectivity),  then the problem can be transformed (for example, by splitting a node with high vertex connectivity into multiple nodes) such that the problem can be reduced to a form that can be mapped to the physically available connectivity. The system is initialized to a random or known initial state. Subsequently, the quantum system is allowed to evolve until it settles down or anneals to a low-energy state as it attempts to minimize the energy function. The resultant low-energy state provides the solution to the problem. Due to the likelihood of the system getting trapped in a local minimum, multiple runs of the algorithm are attempted starting with different initial states, and the lowest energy solution across different runs can be utilized as the solution to the problem.  
 
In general, multiple runs of a quantum algorithm are typically performed for both gate-based computation and annealing-based computations. This is due to the noisy nature of quantum computation, which results from the temperature-dependent entropy in the system. In addition, due to superposition, when the qubits are measured at the end of the computation, low probability states can also occur as solutions, albeit with a reduced likelihood of occurrence, so that the observed solutions post-measurement from the quantum system can vary across different runs. From across these multiple runs, the best solution is chosen by evaluating an energy function for the solutions across different runs, or the solution that occurs more often is chosen as the solution to the problem under consideration.

\subsubsection{Quantum Algorithms}
Quantum computation leverages unique quantum-mechanical properties, such as superposition, entanglement, and interference, to process information encoded in quantum bits (qubits). Unlike classical bits, which are binary, qubits can exist in the superposition of multiple states, enabling quantum algorithms to perform certain computations more efficiently than their classical counterparts. One notable example is Shor's algorithm \cite{qalgo-shor}, which demonstrates a significant speedup in factorizing large integers compared to classical approaches. This algorithm exploits quantum superposition and entanglement to find the prime factors of an integer exponentially faster than the best-known classical algorithms. The implications of Shor's algorithm are profound, particularly in cryptography, as many current encryption methods, such as RSA, rely on the high computational complexity of factorizing large numbers. The advent of quantum computing thus poses a potential threat to these encryption systems. Another prevalent quantum algorithm is Grover's algorithm \cite{qalgo-grover}, designed for searching unstructured databases. Grover's algorithm achieves a quadratic speedup over classical search algorithms. In a database of N unstructured items, Grover's algorithm can find the desired item in roughly $\sqrt{N}$ steps, compared to the N steps required classically. This speedup, while less dramatic than that of Shor's algorithm, is still significant and has implications for a wide range of search-related problems.

In recent years, advancements in quantum algorithms have expanded beyond these foundational concepts, embracing more complex and application-specific algorithms. Among these, the Variational Quantum Eigensolver (VQE) and Quantum Approximate Optimization Algorithm (QAOA) have gained prominence. The Variational Quantum Eigensolver (VQE) \cite{qalgo-vqe} is a hybrid quantum-classical algorithm primarily used in quantum chemistry for finding the ground state energy of molecules. VQE operates by preparing a quantum state on a quantum computer and measuring its energy, while a classical computer variably adjusts the quantum state to minimize its energy. This approach leverages the quantum system for what it does best – representing complex quantum states – and uses classical optimization techniques to navigate the solution space efficiently. The Quantum Approximate Optimization Algorithm (QAOA) \cite{qalgo-qaoa} is designed for solving combinatorial optimization problems, like the Max-Cut problem. QAOA works by encoding the problem into a Hamiltonian and then applying a series of quantum gates that approximate the problem's solution. Its performance improves with the number of quantum layers used, offering a promising approach for achieving quantum advantage in optimization tasks. Quantum Machine Learning (QML) \cite{west_benchmarking_2023, west2023towards} represents another emerging field. QML algorithms leverage quantum computing for machine learning tasks, potentially offering speedups in data processing and pattern recognition. Quantum machine learning is particularly compelling because it merges two cutting-edge fields, quantum computing and artificial intelligence, potentially leading to breakthroughs in both domains.

While these recent quantum algorithms show great promise, they are still in the developmental stage and face implementation challenges. They often require a significant number of qubits and sophisticated error correction methods to be practically viable. However, as quantum technology continues to advance, these algorithms are likely to play a pivotal role in realizing the practical applications of quantum computing.

\subsection{Cloud Computing}
Cloud computing has revolutionized how software and IT infrastructure capabilities are made available as subscription-oriented computing utilities on a pay-as-you-go basis to consumers over the Internet. This paradigm has brought numerous benefits to society, enabling enterprises to expand globally faster and supporting scientific research advancement \cite{buyya_manifesto_2019}. In this section, we briefly discuss their computing models along with the recent emergence of the serverless computing model.

\subsubsection{Cloud computing models}
There are several service models and deployment models in cloud computing:
\begin{enumerate}
    \item \textbf{Service models}: Cloud computing can be delivered through three main service models, including Infrastructure as a Service (IaaS), Platform as a Service (PaaS), and Software as a Service (SaaS) \cite{cloudcomputing-raj}. IaaS provides users access to physical infrastructures like servers, networking, and data storage. IaaS enables users to outsource the infrastructure maintenance and management to a cloud provider and flexibly adjust the resources as needed. PaaS abstracts away the infrastructure and allows developers to think about creating applications on top of a platform that provides different services, such as authentication, database storage, load-balancing, auto-scaling, etc., to provide an environment that enables ease of development. SaaS enables applications to be hosted on a cloud to directly provide services to a user, with the platform and infrastructure aspects abstracted away from the user of the service.
    
    \item \textbf{Deployment models}: Cloud computing can be deployed in different manners, including public clouds, private clouds, and hybrid clouds. For the public cloud, services are provided over the public Internet and are available to anyone willing to pay for them. This model offers scalability, flexibility, and efficiency. In terms of private cloud, infrastructure is dedicated to a single organization, offering more control and security. It can be hosted internally or externally. Hybrid clouds are combinations of public and private clouds, allowing data and applications to be shared between them. This model provides greater flexibility and optimization of existing infrastructure.
\end{enumerate}

\subsubsection{From Serverful to Serverless computing model}
The transition from traditional (or serverful) models to serverless models marks a significant shift in cloud computing. The serverful computing model requires users to manage servers and computation resources, demanding setup, maintenance, and scaling, which offers control but adds complexity \cite{yussupov2021serverful}. Serverless computing, however, abstracts server management, focusing developers on code rather than infrastructure. It dynamically allocates resources, optimizing costs by charging only for actual usage, and automatically scales to demand \cite{serverless-raj}. This shift simplifies operations, reduces costs, and allows for more efficient resource use, emphasizing development efficiency over infrastructure management \cite{shafiei_serverless_2022}. There are two common serverless paradigms: 
\begin{enumerate}
    \item \textbf{Function as a Service (FaaS)}: FaaS allows users to execute code in response to specific events or triggers, typically through an application programming interface (API). The cloud provider automatically allocates and sets up the underlying infrastructure required to run the code \cite{yussupov2021faasten}. Examples of FaaS-based commercial platforms include Amazon Web Services (AWS) Lambda, Google Cloud Functions, and Microsoft Azure Functions. Besides, there are numerous open-source FaaS frameworks, such as OpenFaaS, OpenWhisk, and Knative.
    \item \textbf{Backend as a Service (BaaS)} provides a set of pre-built backend services, such as database management, user authentication, and push notifications, which developers can use to build and run their applications. Examples of popular BaaS platforms include Firebase, AWS Mobile Hub, AWS Amplify, and Microsoft Azure Mobile Apps.
\end{enumerate}

\section{Quantum Cloud Computing and Recent Advances}

\subsection{State-of-the-Art in Quantum Cloud Computing}
\begin{figure}[htbp]
    \centering
    \includegraphics[scale=0.13]{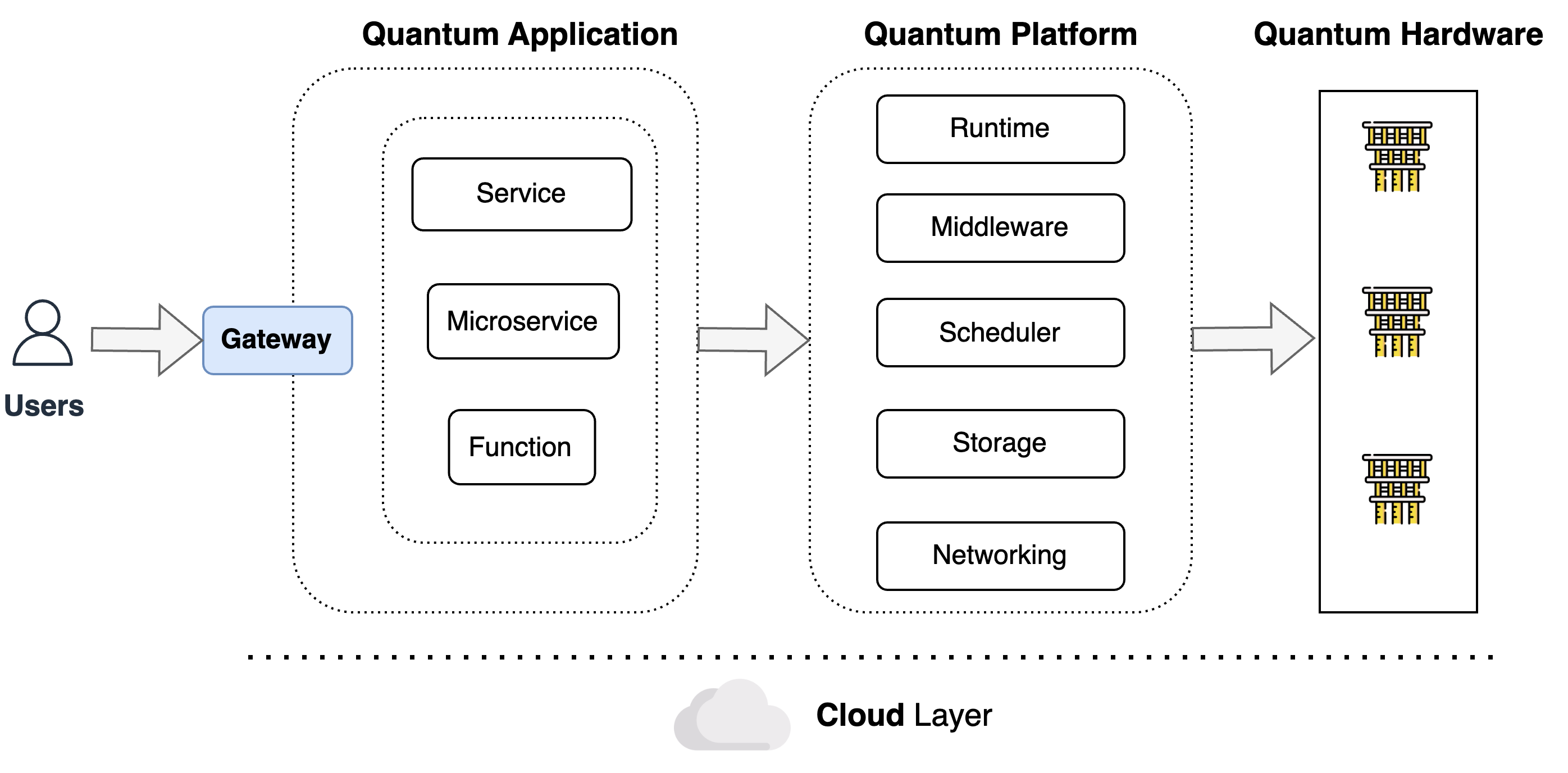}
    \caption{A high-level view of contemporary quantum cloud computing}
    \label{fig:qcloud}
\end{figure}

Quantum Cloud Computing (QCC) represents an emerging computational paradigm that integrates the principles of quantum computing with cloud computing, enabling quantum computations to be executed on cloud-based quantum platforms. This model allows for quantum computational resources to be either publicly accessible via the Internet (public quantum cloud) or privately within a specific organization through a secured network (private quantum cloud), thereby democratizing access to quantum computing capabilities. QCC allows users to perform quantum computing tasks without investing in their own quantum hardware. As depicted in Figure \ref{fig:qcloud}, users interact with these quantum computational resources through a cloud interface, utilizing APIs to access software as a service. This paradigm facilitates the decomposition of large-scale quantum applications into microservices \cite{rojo2021trials} and quantum functions \cite{nguyen2024qfaas}, which can then be efficiently deployed and managed on a cloud-based platform. The cloud platform comprises essential components for orchestrating the quantum runtime and execution environment, resource allocation, storage, and networking. Ultimately, the quantum workload is processed in quantum computers and managed at remote data centers. 

A key innovation in QCC is the potential integration with the quantum Internet. This paradigm promises to revolutionize data communication by utilizing quantum principles for network communications, potentially eliminating the need for classical intermediaries in quantum data exchange. However, the realization of a mature quantum Internet remains a future goal, with current challenges including the development of robust quantum communication protocols and the integration of quantum and classical systems \cite{caleffi_quantum_2018, wehner_quantum_2018}. Presently, the interconnection between cloud-based quantum computers and users still predominantly relies on classical Internet and computing technologies, leading to the emergence of hybrid quantum cloud computing models. These hybrid models combine quantum computational power with classical networking and data processing, offering a pragmatic step towards fully realizing quantum cloud computing's potential.

Quantum cloud computing is still an emerging field; however, it is witnessing increasing interest from both corporate and academic entities focused on developing and deploying these services.  Singh et al. \cite{singh_quantum_2014} highlighted the idea of quantum-cloud integration and implied this combination would be the potential approach of future quantum computing. One of the first trials to bring quantum resources to the cloud was conducted by the Center for Quantum Photonics (CQP) at the University of Bristol \cite{devitt_performing_2016} in early 2016. They introduced a two-qubit optical-based quantum computer, accessible through the Internet for testing purposes, marking the initiation of quantum resources into the cloud domain. Subsequent to this, industry giants such as IBM \cite{qcloud-ibm}, Amazon Web Services (AWS) \cite{qcloud-braket}, and Microsoft Azure \cite{qcloud-azure} began to provide quantum computing services to the public as part of their cloud services offerings. These initiatives underscored the viability and growing accessibility of quantum computing resources via cloud platforms. Karalekas et al. \cite{qcloud-rigetti} detailed the architecture of a typical quantum cloud computing platform, specifying four essential components: 1) an apparatus for accommodating physical qubits, 2) a control system for manipulating the apparatus, 3) an executor for orchestrating the control system, and 4) a compiler for compiling quantum circuit for the executor. Faro et al. \cite{faro2023middleware} introduced a concept for a hybrid quantum cloud computing architecture that incorporates middleware designed to facilitate the integration and management of quantum and classical computations. Furthermore, several works have been proposed to enhance the error robustness of cloud-based quantum computer interfaces. For example, Carvalho et al. \cite{carvalho2021error} achieved substantial error reduction in quantum logic operations using optimized pulses through a cloud quantum computing interface, marking a pivotal advancement in enhancing the reliability and performance of quantum computations. These developments collectively represent the initial steps towards establishing a standardized approach for cloud-based quantum infrastructure across academia and industry, aiming to realize the full potential of quantum cloud computing.

Due to the noisy intermediate-scale nature of available quantum machines \cite{nisq-preskill, 9645257}, high-performance quantum simulators also need to be provided through the cloud to create the experiment environment for quantum-related research. Most quantum cloud providers offer access to large-scale quantum simulators for prototyping quantum applications. For example, IBM Quantum \cite{qcloud-ibm} offers access to various simulators, including the statevector simulator (32 qubits), QASM simulator (32 qubits), and the Clifford simulator, which supports up to 5,000 qubits for the Clifford circuit. Similarly, Amazon Braket and Azure Quantum provide access to numerous quantum simulators, such as state vector simulators, density matrix simulators, and tensor network simulators \footnote{https://docs.aws.amazon.com/braket/latest/developerguide/braket-devices.html}. Additionally, Intel also offers a cloud-ready quantum simulator \cite{guerreschi2020intel} with high-performance capabilities for simulating up to 42-qubit quantum circuits, supporting numerical studies, noise/error modeling, and parallel quantum device emulation. It is also important to note that the number of qubits in a quantum computer is not the only metric for benchmarking its performance. Many factors determine the overall performance of a quantum computer, including the quality of the qubits, the error rate, and the connectivity of the qubits. In \cite{wack_quality_2021}, IBM researchers suggest that scale, quality, and speed are three key parameters to measure the performance of NISQ devices. The number of qubits measures scale, quality is determined by quantum volume \cite{cross2019validating}, and speed is measured by Circuit Layer Operations Per Second (CLOPS). 

Given the significant costs associated with cooling and manufacturing quantum computers \cite{de_leon2021materials, li2021co}, their usage costs can be typically higher than utilizing resources on a classical computer. However, due to the ability to support concurrent superposition of states, quantum computers can naturally execute certain computationally hard tasks significantly faster than a classical computer. It is advantageous to execute such tasks on a quantum computer in such cases where the overall cost to execute the task is significantly lower than on a classical computer. Besides the execution cost, one must consider latency constraints to access a quantum computer. For example,  in a hybrid quantum-classical system, the submission of a task from a classical computer to a quantum computer can involve an access latency of the order of 10ms to 20ms each way, depending on the distance between these computers. Preparing a task for submission to a quantum computer can incur additional latency as well.  A quantum computer's resources are typically shared across different users, so additional considerations, such as a job queuing latency or a task compilation latency \cite{nguyen2024qfaas}, would also need to be considered. In general, one has to carefully consider different constraints, such as the overall cost for execution and the overall latency to execute a task on a quantum computer, before choosing a quantum computer over a classical computer to perform a given task. 

\subsection{Quantum Cloud Computing Models}
\subsubsection{Quantum Computing as a Service (QCaaS)}
Quantum Computing as a Service (QCaaS), also referred to as Quantum as a Service (QaaS), represents a general quantum cloud service model within the evolving quantum cloud computing landscape \cite{nguyen2024qfaas}. This model provides users with remote access to quantum computing resources, facilitating the exploration and application of quantum computing without the need to own and maintain quantum hardware. Stefano et al. \cite{de_stefano_towards_2022} proposed the concept of Quantum-Algorithms-as-a-Service (QAaaS) for hybrid quantum-classical applications. QAaaS abstracts the quantum computing elements from the software development process, enabling developers to focus more on application logic rather than the intricacies of quantum computation. This abstraction is crucial for integrating quantum computing into broader IT infrastructures and for making quantum resources more accessible to a diverse range of users, including those with limited expertise in quantum mechanics. As such, QCaaS is pivotal in democratizing access to quantum computing resources on the cloud, promoting innovation, and accelerating the development of quantum applications across various industries \cite{ahmad2023engineering}.

\subsubsection{Quantum Serverless}
Serverless Quantum Computing is the emerging adoption of a serverless computing model for empowering quantum computing, making it more usable and reliable by abstracting the application deployment and infrastructure setup from users \cite{nguyen2024qfaas}. Following this paradigm, users only need to focus on developing quantum applications using cloud-based services. The serverless applications are highly scalable and effectively utilize the resource, therefore, optimizing the total cost with the pay-per-use model.  The serverless quantum computing model is well-suited to the hybrid quantum-classical application deployment. In 2021, IBM proposed the proof of concept for Quantum Serverless architecture \cite{ibmroadmap}, which allows the incorporation of quantum and classical tasks in a single application. They have claimed that direction is the future of quantum programming and have planned to introduce Quantum Serverless with intelligent orchestration, circuit knitting toolbox, and circuit libraries by 2023 \cite{ibmq-roadmap}.


Cloud-native technologies such as containerization, API gateway, continuous integration, and continuous delivery can be adopted for developing a quantum serverless platform. Garcia-Alonso et al. \cite{q-api-gateway} proposed a proof of concept for Quantum API Gateway (QAPI), which adapts API gateway for exposing quantum services. They also demonstrated that a quantum service could not be deployed permanently inside a quantum computer like its classical counterpart. Instead, the quantum circuit needs to be compiled and sent to an appropriate quantum device for execution at the run time. They also proposed an execution time forecasting model to recommend the best quantum computer for each task and evaluated their proof-of-concept on Amazon Braket service. 
Dreher et al. \cite{dreher_prototype_2019} proposed a simple container-based prototype to encapsulate Qiskit codes into a Docker container on the local computer and interact with cloud-based IBM Quantum service. 
In \cite{rojo2021trials}, the authors demonstrated their trials on deploying hybrid quantum microservices of the Travelling Salesman Problem (TSP) to reveal the limitation of quantum service engineering. They particularly deployed adiabatic and gate-based quantum implementations of TSP on the Amazon Braket platform using IonQ, Rigetti, and D-Wave hardware. The evaluation of this research showed many challenges of quantum service engineering with current NISQ devices, including the limited number of qubits, error rates, response time, and service costs. 
Focusing on the serverless integration aspect, Grossi et al. \cite{qfw-scq-ibm} proposed a Minimum Viable Product (MVP) design to demonstrate the possibility of integrating serverless workflow for quantum computing. This design comprises three architectural layers that involve multiple cloud-native technologies and quantum services from IBM.
However, integrating the services of a single vendor can lead to the data lock-in problem, which is popular for serverless computing. Extending the idea of integrating traditional serverless and cloud-native techniques for quantum computing, QFaaS \cite{nguyen2024qfaas} is one of the first full-stack serverless function-as-a-service platforms that support multiple quantum SDKs and quantum computing services. QFaaS supports the deployment of hybrid quantum-classical applications involving DevOps techniques and a function-as-a-service serverless computing model. The evaluation of QFaaS with practical use cases on multiple quantum software kits and cloud vendors demonstrates the versatility and potential of bringing serverless techniques for accelerating quantum software engineering, which has been adopted in industry platforms \cite{citynow2023enabling}. 

\subsubsection{Hybrid Quantum-Classical Computing}
The current development of almost all quantum cloud computing platforms still depends on classical architecture, generally referred to as the hybrid quantum-classical cloud (HQCC) \cite{qcloud-rigetti}. The sample workflow in an HQCC system is illustrated in Figure \ref{fig:qcloud}. In this architecture, users submit their quantum circuits through web API to the classical cloud services (such as IBM Cloud or Amazon Web Services). These circuits are queued before being forwarded to a quantum processor for execution. Finally, the results will be returned to the classical cloud and forwarded to the original user. All popular quantum cloud providers, such as IBM Quantum, Amazon Braket, and Azure Quantum, are employing this hybrid architecture for their quantum computing service.
HQCC utilizes classical and quantum computing resources in a cloud environment to solve specific problems more efficiently by combining the strengths of both types of computing. The aim is to improve the system's performance by employing the unique capabilities of quantum computing, like quantum parallelism and entanglement, along with the power and flexibility of classical computing. Faro et al. \cite{faro2023middleware} proposed the middleware architecture for orchestrating hybrid quantum-classical computing systems. This heterogeneous architecture highlights the benefits of combining the advantages of both quantum and classical computing to optimize cloud-based resource utilization. Pfandzelter et al. \cite{pfandzelter2023kernel} also proposed a Kernel-as-a-Service (KaaS) programming model to support heterogeneous workflows involving classical and quantum computation tasks. This prototype shows promising development of HQCC architecture design in the future.

\subsection{Quantum Cloud Applications and Use Cases}

Quantum cloud computing is pivotal in the development and deployment of quantum software. As quantum cloud computing is currently the only way to access quantum computers outside developers, almost all empirical studies have been conducted using quantum cloud resources. Table \ref{qcloud-applications} provides several representative works to demonstrate practical use cases of quantum software on the cloud. 

\begin{table}[htbp]
\caption{Representative works on a subset of key use cases of quantum software and services on the quantum cloud environments \\ (QRNG: Quantum Random Number Generation, QML: Quantum Machine Learning)}
\label{qcloud-applications}
\begin{tabular}{|p{0.11\textwidth}|p{0.19\textwidth}|p{0.63\textwidth}|}
\toprule
\textbf{Category} & \textbf{Reference} & \textbf{Highlights} \\
\hline
\multirow{3}{*}{\begin{tabular}[c]{@{}l@{}} QRNG \end{tabular}} & Huang et al. \cite{huang_quantum_2021} & Proposed a cloud-based QRNG on Alibaba Cloud servers by using four different quantum devices (single-photon-detection, phase-fluctuation, photon-counting-detection, and vacuum-fluctuation)  \\ \cline{2-3}
 & Li et al. \cite{li_quantum_2021} & Proposed a source-independent QRNG with parameter optimization and empirically evaluated on IBM Quantum cloud service \\ \cline{2-3}
 & Kumar et al. \cite{kumar_quantum_2022} & Proposed a device-independent quantum random number generator on IBM Quantum based on the combination of multiple Hadamard gates \\
  \hline
\multirow{4}{*}{\begin{tabular}[c]{@{}l@{}} QML \end{tabular}} & Gomez et al. \cite{berganza_gomez_towards_2022}  & Proposed AutoQML, a cloud-based framework to automatically search optimal quantum circuit architecture using qGAN \\ \cline{2-3}
 & Gong et al. \cite{gong_quantum_2021}  & Proposed a quantum k-means algorithm based on trusted quantum cloud servers and quantum homomorphic encryption technique  \\ \cline{2-3}
 & Fadli et al. \cite{Fadli2023-ad}  &  Proposed a hybrid QML model using Quantum Integrated Cloud Architecture (QICA) for aerospace applications and satellite network optimization \\
 \cline{2-3}
 & Hibat-Allah et al. \cite{hibatallah2023framework}  &  Proposed a framework to demonstrate the efficiency of Quantum Circuit Born Machines (QCBMs) in data-limited scenarios for potential practical quantum advantage. \\
  \hline
  
\multirow{2}{*}{\begin{tabular}[c]{@{}l@{}}Chemical \\ Simulation\end{tabular}} & O'Malley et al. \cite{o_malley2016scalable} (Google) & \begin{tabular}[c]{@{}l@{}}Calculated the lowest energy electron arrangement of molecular hydrogen \\  using three qubits\end{tabular} \\ \cline{2-3} 
 & Kandala et al. \cite{kandala2017hardware} (IBM) & \begin{tabular}[c]{@{}l@{}}Calculated the ground-state energy determination using six qubits for \\ molecules up to BeH2\end{tabular} \\ \hline
  
\multirow{3}{*}{\begin{tabular}[c]{@{}l@{}} Other \\ Applications \end{tabular}} & Fang et al. \cite{Fang_2023_crypto}  & Proposed quantum protocols for XOR and AND operations and extending them to multiparty scenarios on quantum cloud \\ \cline{2-3}
 & Gong et al. \cite{Gong2020-as}  & Proposed a quantum homomorphic encryption ciphertext retrieval scheme using Grover's algorithm in quantum cloud computing to enhance data retrieval efficiency  \\ \cline{2-3}

& Azzaoui et al. \cite{el_azzaoui_blockchain-based_2022}  & Proposed a blockchain-based delegated quantum cloud architecture to enhance the security for medical data processing  \\ 
 \bottomrule
\end{tabular}
\end{table}

Quantum Random Number Generation (QRNG), an important application, has seen diverse implementations. For example, Huang et al. \cite{huang_quantum_2021} enhanced cloud cybersecurity by integrating four types of quantum random number generators on Alibaba Cloud servers, combining their outputs for robust random number generation in high-security applications like Alipay. Li et al. \cite{li_quantum_2021} developed a quantum random number generator (QRNG) on IBM's cloud-based quantum computers, addressing the challenge of noise-induced randomness errors. Inspired by source-independent QRNG in optics, their method estimates errors in superposition state preparation, ensuring randomness even with readout errors, and optimizes parameters for increased random bit generation rate. Similarly, Kumar et al. \cite{kumar_quantum_2022} demonstrated a Quantum True Random Number Generator (QTRNG) on IBM's cloud-based quantum computing platform, showcasing a practical application of cloud quantum computing in cryptographic operations.

In terms of quantum machine learning (QML) applications, Gomez et al. \cite{berganza_gomez_towards_2022} developed an Automated QML (AutoQML) framework in a classical-quantum hybrid cloud architecture, enabling parallelized hyperparameter exploration and model training. They demonstrated training a quantum Generative Adversarial neural Network (qGAN) for generating energy prices, showcasing the potential of QML in the energy economics sector. Gong et al. \cite{gong_quantum_2021} proposed a quantum k-means algorithm that employs quantum homomorphic encryption for security, demonstrating its effectiveness in reducing client-side computational burden and protecting data privacy in the cloud. 
Fadli et al. \cite{Fadli2023-ad} presented the Quantum Integrated Cloud Architecture (QICA), aimed at enhancing quantum computing for aerospace applications, especially in satellite networks. They explore the computational benefits of a Hybrid Quantum-Classical machine learning architecture using IBM Quantum and Qiskit Machine Learning, demonstrating QICA's potential in advancing aerospace technology. In their study, Hibat-Allah et al. \cite{hibatallah2023framework} leverage quantum cloud computing to compare the efficacy of classical and quantum generative models, particularly Quantum Circuit Born Machines (QCBMs).

One of the most promising applications of quantum computing lies in simulating molecules and chemical reactions. With quantum computational devices accessible through cloud platforms, researchers in computational chemistry can now perform simulation experiments more efficiently \cite{rossi2021quantum}. Recent advancements from leading technology firms have underscored the potential of quantum cloud computing in molecular simulation. In 2016, a team from Google utilized three qubits to determine the ground-state energy configuration of a hydrogen molecule \cite{o_malley2016scalable}. Following this, IBM Quantum achieved a milestone in 2017 by simulating the properties of three molecules—hydrogen, lithium hydride, and beryllium hydride (BeH2) — using six qubits \cite{kandala2017hardware}. These developments mark significant progress towards accurately predicting properties for novel molecules, a crucial step in the drug discovery process \cite{popkin2017quantum}.

Additionally, there are several applications of quantum cloud computing for cybersecurity and other domains. Fang et al. \cite{Fang_2023_crypto} developed cloud-assisted quantum protocols for enhanced security in applications like Anonymous Voting and Multiparty Private Set Intersection, leveraging quantum cryptography. Gong et al. \cite{Gong2020-as} introduced a novel Quantum Homomorphic Encryption Ciphertext Retrieval (QHECR) scheme based on Grover's algorithm, addressing the challenge of efficiently retrieving homomorphically encrypted data in quantum cloud environments. Azzaoui et al. \cite{el_azzaoui_blockchain-based_2022} proposed a Quantum Cloud-as-a-service for Smart Healthcare, combining Quantum Terminal Machines (QTM) and Blockchain for enhanced security and feasibility. Their architecture offers a scalable and secure solution for complex healthcare computations, highlighting the practicality and robust security of Q-OTP encryption.  In \cite{commercial-apps-qt}, Bova et al. illustrated the vision of the commercial application of quantum computing cybersecurity, materials and pharmaceuticals, banking, and finance. Similarly, Hassija et al. \cite{hassija2020forthcoming} highlighted potential applications of quantum computing in various practical scenarios such as logistics, financial risk analysis, and satellite communication. These examples showcase the versatility and potential of quantum cloud services in various domains.

\subsection{Quantum Cloud Providers and Platforms}
\subsubsection{Cloud-enabled quantum hardware providers}
Most quantum hardware vendors offer cloud-based access to their computation resources. We summarize the latest developments of popular cloud-enabled quantum hardware vendors in Table \ref{tab:quantum-vendors}. Several vendors, such as IBM Quantum \cite{qcloud-ibm}, Rigetti \cite{qcloud-rigetti}, and IonQ \cite{qcloud-ionq}, offer cloud-based computing services to their quantum computers, while other providers like Amazon Braket \cite{qcloud-braket} and Azure Quantum \cite{qcloud-azure} mainly collaborate with third-party hardware vendors to offer quantum computing services. Various technologies such as superconducting, trapped ions, neutral atoms, and quantum annealing have been employed in building quantum chips (or quantum processing units - QPUs). Each technique leverages distinct physical systems to implement qubits and quantum operations \cite{de_leon2021materials}.

\begin{table}[htbp]
\caption{A summary of quantum hardware vendors with cloud-based access until March 2024. \\ (* Annealing qubit; **: Number of variables (qudit); \checkmark: Yes, -: Not available)}
\label{tab:quantum-vendors}
\begin{tabular}{|l|l|l|ll|l|ll|}
\hline
\multirow{2}{*}{\textbf{Hardware vendor}} & \multirow{2}{*}{\textbf{\begin{tabular}[c]{@{}l@{}}QPU \\ technology\end{tabular}}} & \multirow{2}{*}{\textbf{\begin{tabular}[c]{@{}l@{}}Max \\ qubit \\ count\end{tabular}}} & \multicolumn{2}{l|}{\textbf{Qubit topology}} & \multirow{2}{*}{\textbf{\begin{tabular}[c]{@{}l@{}}Supported \\ SDKs\end{tabular}}} & \multicolumn{2}{l|}{\textbf{Cloud-based access}} \\ \cline{4-5} \cline{7-8} 
 &  &  & \multicolumn{1}{l|}{\begin{tabular}[c]{@{}l@{}}Partially\\ connected\end{tabular}} & \begin{tabular}[c]{@{}l@{}}Fully\\ connected\end{tabular} &  & \multicolumn{1}{l|}{\begin{tabular}[c]{@{}l@{}}Vendor's \\ Cloud\end{tabular}} & \begin{tabular}[c]{@{}l@{}}External \\ Cloud\end{tabular} \\ \hline
\begin{tabular}[c]{@{}l@{}}Alpine Quantum \\ Technologies \cite{qcloud-aqt} \end{tabular} & \begin{tabular}[c]{@{}l@{}}Trapped \\ Ion\end{tabular} & 20 & \multicolumn{1}{l|}{} & \checkmark & \begin{tabular}[c]{@{}l@{}}Qiskit, \\ Pennylane\end{tabular} & \multicolumn{1}{l|}{\checkmark} & - \\ \hline
Atom Computing \cite{qcloud-atom} & \begin{tabular}[c]{@{}l@{}}Neutral Atom\end{tabular} & 1180 & \multicolumn{1}{l|}{\checkmark} &  & - & \multicolumn{1}{l|}{-} & - \\ \hline
D-Wave \cite{qcloud-dwave} & \begin{tabular}[c]{@{}l@{}}Quantum \\ Annealing\end{tabular} & 5000* & \multicolumn{1}{l|}{\checkmark} &  & Ocean & \multicolumn{1}{l|}{\checkmark} & \checkmark \\ \hline
Google \cite{qcloud-googleqc} & \begin{tabular}[c]{@{}l@{}}Super-\\ conducting\end{tabular} & 54 & \multicolumn{1}{l|}{\checkmark} &  & Cirq & \multicolumn{1}{l|}{\checkmark} & - \\ \hline
IBM Quantum \cite{qcloud-ibm} & \begin{tabular}[c]{@{}l@{}}Super-\\ conducting\end{tabular} & 1121 & \multicolumn{1}{l|}{\checkmark} &  & Qiskit & \multicolumn{1}{l|}{\checkmark} & \checkmark \\ \hline
IonQ \cite{qcloud-ionq} & Trapped Ion & 36 & \multicolumn{1}{l|}{} & \checkmark & \begin{tabular}[c]{@{}l@{}}Qiskit, Cirq, tket,\\ Q\#, Pennylane,\\ Forge, ProjectQ\end{tabular} & \multicolumn{1}{l|}{\checkmark} & \checkmark \\ \hline
\begin{tabular}[c]{@{}l@{}}Oxford Quantum\\ Circuits (OCQ) \cite{qcloud-ocq}\end{tabular} & \begin{tabular}[c]{@{}l@{}}Super-\\ conducting\end{tabular} & 32 & \multicolumn{1}{l|}{\checkmark} &  & tket & \multicolumn{1}{l|}{\checkmark} & \checkmark \\ \hline
Pasqal \cite{qcloud-pasqal}& \begin{tabular}[c]{@{}l@{}}Neutral Atom\end{tabular} & 100 & \multicolumn{1}{l|}{\checkmark} &  & Pulser & \multicolumn{1}{l|}{\checkmark} & \checkmark \\ \hline
\begin{tabular}[c]{@{}l@{}}Quantum Computing \\ Inc (QCI) \cite{qcloud-qci} \end{tabular} & \begin{tabular}[c]{@{}l@{}}Photonics \\ (Entropy)\end{tabular} & 949** & \multicolumn{1}{l|}{ } & \checkmark & Qatalyst & \multicolumn{1}{l|}{\checkmark} & - \\ \hline
Quantinuum \cite{qcloud-quantinuum}& \begin{tabular}[c]{@{}l@{}}Trapped \\ Ion\end{tabular} & 32 & \multicolumn{1}{l|}{} & \checkmark & tket, lambeq & \multicolumn{1}{l|}{-} & \checkmark \\ \hline
QuEra \cite{qcloud-quera} & \begin{tabular}[c]{@{}l@{}}Neutral Atom\end{tabular} & 256 & \multicolumn{1}{l|}{\checkmark} &  & Braket & \multicolumn{1}{l|}{-} & \checkmark \\ \hline
\begin{tabular}[c]{@{}l@{}} Quantum Inspire \\ (QuTech) \cite{qcloud-qinspire}\end{tabular} & \begin{tabular}[c]{@{}l@{}}Super- \\ conducting, \\ Solid-state spins\end{tabular} & 5 & \multicolumn{1}{l|}{\checkmark} &  & cQASM, QI & \multicolumn{1}{l|}{\checkmark} & \checkmark \\ \hline  
Rigetti \cite{qcloud-rigetti} & \begin{tabular}[c]{@{}l@{}}Super-\\ conducting\end{tabular} & 84 & \multicolumn{1}{l|}{\checkmark} &  & Quil & \multicolumn{1}{l|}{\checkmark} & \checkmark \\ \hline
Xanadu  \cite{qcloud-xanadu} & \begin{tabular}[c]{@{}l@{}}Photonics\end{tabular} & 24 & \multicolumn{1}{l|}{\checkmark} &  & \begin{tabular}[c]{@{}l@{}}Pennylane, \\ Strawberry Fields\end{tabular} & \multicolumn{1}{l|}{\checkmark} & - \\ \hline
\end{tabular}
\end{table}

The superconducting technique leverages the properties of superconducting circuits to create qubits. This approach uses Josephson junctions to exploit quantum mechanical phenomena at macroscopic scales, enabling the manipulation of quantum states. Superconducting qubits are known for their relatively more straightforward integration into electronic systems and potential for scalability \cite{krantz2019quantum}. This approach can be considered the most popular technique for developing quantum computers by many leading companies such as IBM Quantum \cite{qcloud-ibm}, Rigetti \cite{qcloud-rigetti}, Oxford Quantum Circuits (OCQ) \cite{qcloud-ocq}, and Google \cite{qcloud-googleqc}. In 2019, Google claimed quantum supremacy with their Sycamore 54-qubit superconducting processor, demonstrating it could perform a specific task in 200 seconds that would take the best classical supercomputer approximately 10,000 years, marking a significant breakthrough in quantum computing \cite{arute2019quantum}. However, access to Google's quantum processor has remained limited to the general public compared to other hardware vendors. IBM Quantum offers a range of quantum processors available on their cloud-based platform, from 5 qubits to 433 qubits. In 2023, they released the Condor processor with 1,121 superconducting qubits and a high-performance 133-qubit Heron processor, marking a significant milestone in quantum hardware development \cite{ibmq-1121qubit}. IBM also demonstrated the utility in quantum computing with their 127-qubit superconducting processor, showcasing accurate expectation values measurement beyond classical computation capabilities, highlighting advancements in coherence, calibration, and noise management as critical enablers for pre-fault-tolerant quantum computing applications \cite{kim2023evidence}. Other companies such as Rigetti \cite{qcloud-rigetti}, Oxford Quantum Circuits (OCQ) \cite{qcloud-ocq}, and Quantum Inspire (QuTech) \cite{qcloud-qinspire} also employed superconducting techniques for developing their cloud-based quantum hardware. QuTech also demonstrates their second hardware technique to develop a 2-qubit quantum processor based on single electron spin qubits in silicon \cite{pla2012single, vahapoglu2022coherent}.

Another popular quantum hardware technique is trapped ion, which leverages ions (charged atoms) trapped in electromagnetic fields to function as qubits. Lasers are used to perform qubit initialization, manipulation, and measurement. This method is known for its long coherence times and high-fidelity operations. IonQ \cite{qcloud-ionq}, Alpine Quantum
Technologies (AQT) \cite{qcloud-aqt}, and Honeywell Quantum (now part of Quantinuum) \cite{qcloud-quantinuum} are prominent vendors in this area, with all fully-connected qubit devices developed, demonstrating advanced quantum computing systems based on trapped ions. Besides, neutral atom technology, which uses lasers to trap and cool neutral atoms (atoms with no net electric charge) in a 2D or 3D array, also caught attention. This technology promises scalability and high qubit numbers, as Atom Computing \cite{qcloud-atom} claimed that they successfully developed a 1,180-qubit quantum computer in 2023 and are preparing to make their system available for cloud-based access. Pasqal \cite{qcloud-pasqal} and QuEra \cite{qcloud-quera} are other notable examples of companies exploring quantum computing with neutral atoms to develop their cloud-based hardware. Furthermore, photonics is also a promising technique for developing quantum chips \cite{slussarenko2019photonic}. Quantum Computing Inc (QCI) introduced their Dirac-3 quantum system as a new approach in quantum computing through Entropy Quantum Computing (EQC) \cite{qcloud-qci} to support higher-order interactions among qudits and leveraging entropy and noise, enabling operation at room temperature without cryogenic environment, and promising enhanced computing speed and capacity. Utilizing photonics, QCI's system advances beyond traditional two-level qubits by implementing multi-level qudits \cite{campbell2014enhanced}, with every single photon enabling up to 10,000 levels through various degrees of freedom, like polarization and orbital angular momentum \cite{ringbauer2022universal}. This capability allows for the management of up to 949 photons/qudits, where each photon's degrees of freedom are harnessed to represent multiple levels, effectively turning each variable into a qudit with its level signifying the variable's value \cite{qcloud-qci}. Research into EQC is underway, with the potential to significantly enhance the energy efficiency of quantum cloud computing by minimizing the cooling requirements of quantum computational systems. Another company, Xanadu, \cite{qcloud-xanadu} has developed its nanophotonic quantum chip that can operate at room temperature, offering advantages in stability and scalability \cite{arrazola2021quantum}.

While superconducting, trapped ion, photonics, and neutral atom techniques are employed to develop gate-based quantum computers, the quantum annealing approach is designed to solve optimization problems by naturally finding a quantum system's ground state. It uses a quantum mechanical process to minimize energy states and find solutions to optimization problems. D-Wave Systems \cite{qcloud-dwave} is the most well-known company specializing in quantum annealing technology, offering quantum cloud computing services for specific optimization tasks. However, they recently added the gate-based quantum model to their roadmap towards developing devices for universal quantum computation. 

The selection of quantum hardware technology plays a pivotal role in determining the efficiency and type of computations that can be executed, directly affecting quantum computing systems' scalability, coherence time, and error rates. As this field progresses, the refinement and interaction of these technologies will be fundamental in developing practical and broadly accessible quantum cloud computing services. Furthermore, the cloud-based provision of quantum computational resources simplifies the integration with high-performance computing (HPC) resources, promising enhanced computational capabilities and broader application potential.

\subsubsection{Quantum cloud computing services and platforms}
Apart from cloud services offered directly from the hardware vendors, as discussed in the previous section, we summarize other cloud services and frameworks for quantum computing in Table \ref{tab:quantum-frameworks}.
Most cloud-based quantum computing platforms provide commercial quantum computing as a service. Major well-known cloud platforms such as Amazon Web Services (AWS), Microsoft Azure, Google Cloud, and IBM Cloud have started offering quantum computing services in their cloud ecosystem, opening opportunities for integrating quantum computation tasks with classical ones. IBM offers access to its self-developed quantum computation resources via IBM Cloud \cite{qcloud-ibmcloud} and IBM Quantum platform \cite{qcloud-ibm}, while AWS and Azure provide cloud services and platforms to access other quantum hardware vendors, such as Rigetti, IonQ, and Quantinuum, OCQ, Pasqal, and QuEra. These quantum cloud platforms also offer a range of software development kits (SDKs) for designing, developing, and evaluating quantum applications. For example, IBM Quantum provides users with multiple tools, such as IBM Quantum Composer and IBM Quantum Lab, for designing, visualizing, executing, and analyzing quantum applications. Amazon Braket, Azure Quantum, and Google Cloud also offer their quantum SDKs, naming Braket, Microsoft Quantum SDKs (along with Q\# programming language), and Cirq, respectively. Strangeworks collaborates with almost all major quantum computing hardware and software vendors to create a comprehensive cloud platform that offers users access to nearly all available quantum hardware and other software platforms. 

\begin{table}[htbp]
\caption{Representative quantum cloud platforms and frameworks, until March 2024 \\ (CQS: Commercial quantum service, OSF: Open-source framework, QSim: Quantum simulator, \checkmark: Yes)}
\label{tab:quantum-frameworks}
\begin{tabular}{|l|l|l|l|l|lll|}
\hline
\multicolumn{1}{|c|}{\multirow{2}{*}{\textbf{\begin{tabular}[c]{@{}c@{}}Platforms/ \\ Framework\end{tabular}}}} & \multicolumn{1}{c|}{\multirow{2}{*}{\textbf{Type}}} & \multicolumn{1}{c|}{\multirow{2}{*}{\textbf{\begin{tabular}[c]{@{}c@{}}Quantum \\ Backends\end{tabular}}}} & \multicolumn{1}{c|}{\multirow{2}{*}{\textbf{\begin{tabular}[c]{@{}c@{}}Supported\\  SDKs\end{tabular}}}} & \multirow{2}{*}{\textbf{\begin{tabular}[c]{@{}l@{}}Serverless\\ Model\end{tabular}}} & \multicolumn{3}{l|}{\textbf{Quantum Computing Models}} \\ \cline{6-8} 
\multicolumn{1}{|c|}{} & \multicolumn{1}{c|}{} & \multicolumn{1}{c|}{} & \multicolumn{1}{c|}{} &  & \multicolumn{1}{l|}{Simulation} & \multicolumn{1}{l|}{Annealing} & Gate-based \\ \hline
1Qloud \cite{qcloud-1qloud} & CQS & 1Qbit & 1Qbit &  & \multicolumn{1}{l|}{\checkmark} & \multicolumn{1}{l|}{} &  \\ \hline
\begin{tabular}[c]{@{}l@{}}Amazon \\ Braket \cite{qcloud-braket}\end{tabular} & CQS & \begin{tabular}[c]{@{}l@{}}Rigetti, OCQ,\\ IonQ, QuEra\end{tabular} & \begin{tabular}[c]{@{}l@{}}Braket, Qiskit,\\ Pennylane\end{tabular} & \checkmark & \multicolumn{1}{l|}{\checkmark} & \multicolumn{1}{l|}{} & \checkmark \\ \hline
\begin{tabular}[c]{@{}l@{}}Azure \\ Quantum \cite{qcloud-azure}\end{tabular} & CQS & \begin{tabular}[c]{@{}l@{}}Quantinuum,\\ Rigetti, IonQ, \\ Pasqal\end{tabular} & \begin{tabular}[c]{@{}l@{}}Q\#, Qiskit, \\ Cirq,\end{tabular} &  & \multicolumn{1}{l|}{\checkmark} & \multicolumn{1}{l|}{} & \checkmark \\ \hline
\begin{tabular}[c]{@{}l@{}}Google \\ Cloud \cite{qcloud-googleqc}\end{tabular} & CQS & Google, IonQ  & Cirq &  & \multicolumn{1}{l|}{\checkmark} & \multicolumn{1}{l|}{} & \checkmark \\ \hline
\begin{tabular}[c]{@{}l@{}}IBM \\ Cloud \cite{qcloud-ibmcloud}\end{tabular} & CQS & IBM Quantum & Qiskit & \checkmark & \multicolumn{1}{l|}{\checkmark} & \multicolumn{1}{l|}{} & \checkmark \\ \hline
PlanQK \cite{planqk} & CQS & \begin{tabular}[c]{@{}l@{}}IBM Quantum,\\ Amazon Braket, \\ Azure Quantum\end{tabular} & \begin{tabular}[c]{@{}l@{}}Qiskit, \\ Pennylane\end{tabular} & \checkmark & \multicolumn{1}{l|}{\checkmark} & \multicolumn{1}{l|}{\checkmark} & \checkmark \\ \hline
QEMIST \cite{qcloud-qemist} & CQS & 1Qbit & OpenQEMIST &  & \multicolumn{1}{l|}{\checkmark} & \multicolumn{1}{l|}{} &  \\ \hline
QFaaS \cite{nguyen2024qfaas} & OSF & \begin{tabular}[c]{@{}l@{}}IBM Quantum, \\ Strangeworks\end{tabular} & \begin{tabular}[c]{@{}l@{}}Qiskit, Cirq, \\ Q\#\end{tabular} & \checkmark & \multicolumn{1}{l|}{\checkmark} & \multicolumn{1}{l|}{} & \checkmark \\ \hline
\begin{tabular}[c]{@{}l@{}}QuantumPath \\ \cite{qfw-quantumpath}\end{tabular} & CQS & \begin{tabular}[c]{@{}l@{}}IBM Quantum, \\ Amazon Braket,\\ D-Wave, QuTech\end{tabular} & \begin{tabular}[c]{@{}l@{}}Qiskit, Ocean,\\ Braket, Q\#\end{tabular} &  & \multicolumn{1}{l|}{\checkmark} & \multicolumn{1}{l|}{\checkmark} & \checkmark \\ \hline
\begin{tabular}[c]{@{}l@{}}Strangeworks\\ \cite{qfw-strangeworks}\end{tabular} & CQS & \begin{tabular}[c]{@{}l@{}}IBM Quantum, \\ Amazon Braket,\\ Azure Quantum, \\ Hitachi, Toshiba,...\end{tabular} & \begin{tabular}[c]{@{}l@{}}Qiskit, Braket, \\ Rigetti\end{tabular} & \checkmark & \multicolumn{1}{l|}{\checkmark} & \multicolumn{1}{l|}{\checkmark} & \checkmark \\ \hline
\end{tabular}
\end{table}

Besides, several cloud-based quantum software-oriented frameworks have been proposed. Hevia et al. \cite{qfw-quantumpath} proposed the QuantumPath platform to support the development of quantum applications. This framework integrates multiple visual editors to aid the design of quantum circuits and supports different SDKs with the tendency to become an agnostic software framework for quantum computing. QFaaS \cite{nguyen2024qfaas} is another open-source quantum software framework that supports developing serverless quantum function-as-a-service applications. QFaaS incorporates multiple quantum SDKs and programming languages and can be extendable to work with different quantum cloud providers, such as IBM Quantum and Amazon Braket. Furthermore, PlanQK \cite{planqk} is a comprehensive quantum platform and ecosystem to support the development of quantum workflow applications with a vision towards establishing a quantum application marketplace. Besides, several orchestration tools, such as Orquestra \cite{zapataOrquestraPlatform} and QuantMe \cite{Weder2020-wn}, are proposed for integrating classical components and workflow with quantum algorithms. Xin et al. \cite{xin2018nmrcloudq} also introduced their quantum cloud computing service, named NMRCloudQ, in which quantum hardware is developed based on a nuclear magnetic resonance (NMR) spectrometer. Apart from offering quantum computing services, most quantum cloud platforms and frameworks provide simulation environments for prototyping and developing quantum applications. Several cloud platforms, such as 1QCloud \cite{qcloud-1qloud} and QEMIST \cite{qcloud-qemist}, are dedicated to providing quantum simulation services for specific problems such as optimization and quantum chemistry.

\subsection{Quantum Cloud Resource Management}
Similar to traditional cloud computing, quantum resources must be efficiently managed and allocated as quantum cloud computing continuously advances. However, quantum cloud resource management is facing more complicated challenges.
The first challenges arise due to the heterogeneity of quantum resources in terms of qubit numbers, qubit connectivity, error rates, and quantum processor speed. Indeed, each quantum hardware technology has different advantages and limitations in its performance and scalability. For example, trapped ion quantum devices can achieve higher quantum volume (qubit quality) but face difficulty achieving high processing speed (CLOPS) \cite{blinov_comparison_2021, wack_quality_2021}. In contrast, the opposite can be seen in spin-based quantum devices. Second, quantum resources are almost fixed, i.e., they cannot be divided and scaled flexibly in the same ways as classical ones. We can create multiple virtual machines or containers inside a single classical host machine or a cluster of machines as isolated environments for executing independent tasks concurrently. However, no corresponding techniques for quantum resources have been proposed yet. Besides, each quantum task has different requirements, which are unknown or unpredictable, to allocate an appropriate quantum device for execution. The uncertainty of quantum task requirements and the fixed resource amount of quantum computers accelerate the complexity of quantum resource management problems and result in underestimated or overestimated resource allocation.
Therefore, two key sub-problems of quantum resource management must be considered: resource estimation of quantum tasks and resource allocation (or job scheduling).

\begin{table}[htbp]
\caption{Representative works in quantum cloud resource management from both theoretical and empirical perspective}
\label{qcloud-rm}
\begin{tabular}{|p{0.1\textwidth}|p{0.19\textwidth}|p{0.63\textwidth}|}
\toprule
\textbf{Category} & \textbf{Reference} & \textbf{Highlights} \\
\hline
\multirow{3}{*}{\begin{tabular}[c]{@{}l@{}} Resource \\Estimation\end{tabular}} & Suracha et al. \cite{suchara_qure_2013} & Propose QuRE for quantum resource estimation, enhancing design efficiency through error rate analysis and various error correction strategies.  \\ \cline{2-3} 
 & JavadiAbhari et al. \cite{javadiabhari_scaffcc_2014} & Propose ScaffCC, a scalable framework for quantum circuit analysis and resource estimation, optimizing hardware-agnostic implementations. \\ \cline{2-3}
 & Salm et al. \cite{salm_nisq_2020} & Propose NISQ Analyzer for optimizing quantum backend selection based on task-specific requirements and available resources.  \\
 \hline
\multirow{8}{*}{\begin{tabular}[c]{@{}l@{}} Resource \\Allocation \\ \& Task \\Scheduling\end{tabular}} & Ravi et al. \cite{ravi2021quantum} & Propose the first statistical-based resource management strategy for quantum job using IBM Quantum \\ \cline{2-3}
 & Ngoenriang et al. \cite{ngoenriang_optimal_2022} & Propose a two-stage stochastic resource allocation technique for distributed quantum computing \\ \cline{2-3}
 & Kaewpuang et al. \cite{kaewpuang_stochastic_2022} & Propose a stochastic qubit allocation for quantum cloud under multiple uncertainties of circuit characteristics and waiting time \\ \cline{2-3}
  & Cicconetti et al. \cite{cicconetti_resource_2022} & Propose a network resource allocation for distributed quantum computing base on Weighted Round Robin algorithm \\ \cline{2-3}
& Zhang et al. \cite{10.1145/3555962.3555964} & Proposed a task scheduling method to classify users and tasks and provide differentiated service opportunities 
\\ \cline{2-3}
& Weder et al. \cite{weder2021automated} & Proposed an automated quantum hardware selection for quantum workflows based on quantum circuit properties
\\ \cline{2-3}
& Liu et al. \cite{liu_qucloud_2021} & Propose QuCloud for efficient multi-programming in quantum cloud computing, enhancing resource utilization and reducing error rates\\ \cline{2-3} 
 & Liu et al. \cite{qucloudplus} & Extend QuCloud with QuCloud+, incorporating advanced quantum program profiling to improve qubit fidelity and reduce SWAP overhead. \\

\hline
\multirow{2}{*}{\begin{tabular}[c]{@{}l@{}} Modeling \& \\ Simulation \end{tabular}} & Nguyen et al. \cite{nguyeniquantum2024} & Propose iQuantum toolkit for support the modeling and simulation of quantum resource management in cloud-based environments\\ 
 \bottomrule
\end{tabular}
\end{table}

\subsubsection{Quantum Resource Estimation}
Suracha et al. \cite{suchara_qure_2013} proposed QuRE, a Quantum Resource Estimator to quantify quantum computing resources, including qubit count, execution duration, success probabilities, and gate operations, tailored to specific quantum tasks. They extended their analysis to incorporate diverse error correction strategies, assessing how variations in error rates influence the overall resource estimations, thereby enabling more efficient quantum computing designs. JavadiAbhari et al. \cite{javadiabhari_scaffcc_2014} developed ScaffCC, a comprehensive framework designed for the compilation and analysis of quantum circuits, focusing on scalability and the estimation of resources such as circuit width, depth, gate count, and qubit interactions. This framework enhances hardware-agnostic quantum circuit implementation, optimizing for both small and large-scale applications by addressing the critical aspects of quantum circuit complexity and resource allocation. Salm et al. \cite{salm_nisq_2020} presented the NISQ Analyser as a novel tool designed to optimize the quantum backend selection process by evaluating the specific requirements of quantum tasks. It assesses essential parameters, including qubit requirements, circuit depth, and gate complexity, to ensure the most efficient alignment of quantum algorithms with available quantum computing resources, thereby enhancing the practicality of quantum computing applications.

\subsubsection{Quantum Resource Allocation and Task Scheduling}
When designing a comprehensive resource allocation for the quantum cloud computing paradigm, multiple factors must be considered. In \cite{ravi2021quantum}, Ravi et al. analyzed multiple quantum jobs and resource utilization characteristics using IBM Quantum Cloud services for two years. They evaluated the significance of the execution times, waiting times, and compilation times of quantum circuits, fidelity, error rates, and the utilization of quantum machines. This study gave insights into numerous factors affecting quantum task execution on quantum cloud devices. In the following study \cite{ravi_adaptive_2021}, they proposed a prediction model to predict the fidelity of quantum computers and queueing time for each device based on historical data of IBM Quantum computers. To schedule incoming jobs to an appropriate quantum computer, this work compiles and transpiles the corresponding quantum circuit for each quantum computer, extracts the key features of the circuit, and maps it with the characteristics of the machine. Then, using correlation coefficient techniques, the best-suited quantum computer will be selected for job execution. Although the proposed method, which uses statistical analysis, is straightforward, this work can be considered the first paper on resource management for quantum cloud computing.
Ngoenriang et al. \cite{ngoenriang_optimal_2022} proposed a two-stage stochastic programming-based resource allocation technique for distributed quantum computing, which aims to minimize the total deployment cost while maximizing the utilization of quantum resources. This paper considered multiple uncertainties of the distributed quantum computing model, including quantum task demands, computation power, and the degradation in qubit fidelity through quantum networks. 
Similarly, Kaewpuang et al. \cite{kaewpuang_stochastic_2022} suggested another two-stage stochastic qubit allocation for quantum cloud under the uncertainties of quantum circuit requirements and the expected waiting time. In this study, resource allocation is determined using historical data in the reservation stage and the actual requirements in the on-demand stage. Focusing on the quantum network aspects, Cicconetti et al. \cite{cicconetti_resource_2022} presented another resource allocation technique for distributed quantum computing. They leveraged the Weighted Round Robin (WRR) algorithm to design the network resource allocation technique for quantum applications. The key idea is to pre-calculate the weight of all traffic flows in each quantum application and use the round-robin strategy to assign network resources. They also proposed a quantum network provisioning simulator for the evaluation and showed the trade-offs between fairness and time complexity of the network resource allocation algorithm.
Zhang et al. \cite{10.1145/3555962.3555964} propose a new task scheduling scheme for cloud-based quantum computing platforms, focusing on user and task classification. This approach aims to reduce waiting times for high-priority users, thereby enhancing their experience and addressing the challenge of limited access to high-quality quantum computing resources.

Due to the error-prone nature of NISQ devices, only quantum circuits with a limited number of qubits can be precisely executed. It leads to the common problem of under-utilizing quantum resources as only more qubits are wasted, and only one circuit can be executed each time. Therefore, parallel processing and multi-programming techniques are essential for maximizing the resource utilization for quantum cloud computing \cite{niu_how_2022}.
Das et al. \cite{das_case_2019} presented their studies on quantum multi-programming and proposed several solutions for enabling the multi-programmed NISQ devices. Specifically, they proposed three methods, including 1) a qubit partitioning method to ensure fairness in qubit allocation, 2) a Delayed Instruction Scheduling policy to reduce the multi-program interference, and 3) an Adaptive Multi-Programming to allow flexible switching between single- and multi-programming.
Ohkura et al. \cite{ohkura_simultaneous_2022} proposed \textit{palloq}, a parallel allocation protocol for quantum circuits, which accelerates the performance of quantum multi-programming in NISQ devices. They also considered error detection using randomized benchmarking methods. Nguyen et al. \cite{nguyen_software_2022} developed a full-stack software framework to enable parallel quantum computing for hybrid quantum workloads. This framework supports three modes to distribute quantum tasks, including Message Passing Interface (MPI) protocol and local and cloud-based quantum accelerators, which are built based on nitrogen-vacancy (NV) centers in diamond at Quantum Brilliance company. Weder et al. \cite{weder2021automated} proposed an automated quantum hardware selection approach for quantum workflows, addressing the complexity of selecting suitable quantum hardware based on input data and quantum circuit properties. This approach involves transforming QuantME workflow models, which are used for orchestrating various tasks such as pre-processing, quantum circuit execution, and post-processing tasks of one or multiple quantum algorithms, into native workflow models to select quantum hardware during workflow runtime dynamically, considering the limitations of today's quantum computers. In practice, IBM Quantum uses fair-share scheduling algorithm\footnote{https://quantum-computing.ibm.com/lab/docs/iql/manage/systems/queue/} to ensure fairness in their Open plan quantum cloud service, which is freely public for the research community. When a new quantum job arrives, it is placed in the waiting queue from all users, and its order to be executed will be dynamically determined using the fair-share algorithm. 

In the domain of quantum cloud resource management, strategic qubit allocation and circuit transpilation are pivotal. Liu et al. proposed QuCloud \cite{liu_qucloud_2021}, a qubit mapping technique to optimize the utilization of quantum resources while mitigating error rates and fidelity degradation, which is essential for cloud-based quantum computing. By partitioning physical qubit topologies and introducing the X-SWAP design to minimize SWAP overheads, QuCloud enhances inter-program interactions. Subsequently, the authors proposed QuCloud+ \cite{qucloudplus} to refine this approach further, leveraging advanced quantum program profiling to improve qubit fidelity and reduce SWAP overhead. This development is particularly notable for its support of both 2D and 3D quantum chips, alongside its flexibility in switching between single- and multiple-program mappings on demand. Such developments are crucial for effectively managing and utilizing quantum cloud resources, aligning with the broader objectives of maximizing computational efficiency and adaptability in quantum cloud platforms.

\subsubsection{Simulation Toolkits for Quantum Cloud Resource Management Problems}

In the evolving landscape of quantum cloud computing, where accessibility to physical quantum resources is limited, simulation toolkits like CloudSim \cite{cloudsim} in the classical domain are pivotal for designing and evaluating quantum cloud resource management algorithms. Nguyen et al. proposed iQuantum \cite{iquantum, nguyeniquantum2024}, a toolkit for modeling and simulating resource scheduling algorithms on the cloud. This toolkit is particularly beneficial for simulating quantum computing scenarios that integrate cloud-based quantum resources, focusing on job scheduling and hybrid quantum-classical task orchestration. Furthermore, Passian et al. \cite{passian2022concept} proposed a simulator for quantum edge computing in the coming years. This concept highlights the importance of standardization and the creation of such simulators and toolkits for cloud-based quantum computing, as well as the extension to edge computing environments.

\subsection{Quantum Cloud Security and Privacy}
As quantum cloud computing is in its infancy, a limited number of works in the literature focus on its security and privacy. On the contrary, many studies pay extensive attention to quantum security \cite{chen_quantum_2021, saki_survey_2021,lipinska_secure_2020, wallden_cyber_2019}, post-quantum cryptography \cite{bernstein_post-quantum_2017, ryan_post-quantum_2016}, and quantum-safe techniques \cite{iftikhar_quantum_2021}.

\subsubsection{Security and data privacy threats of quantum cloud computing}
In quantum cloud computing, the integration of quantum services via cloud platforms can introduce several security and privacy challenges. The accessibility of cloud-based quantum computing services potentially enables adversaries to exploit these resources for unauthorized access to sensitive data without needing proprietary quantum hardware. Initial attack vectors can target the exploitation of quantum resources to compromise non-quantum-safe infrastructures or the theft of credentials securing cloud-based quantum services, enabling service manipulation or compromise \cite{hrda2023confidential}. The necessity for secure data transfer channels in remote quantum computing emphasizes the importance of network authenticity, especially in cloud environments where physical network characteristics are abstracted \cite{ghosh2023primer}.

Malicious third-party cloud computers and the vulnerabilities they introduce, such as hacking, data leaks, and insecure APIs, are well-known threats in the classical domain \cite{singh2017cloud}. As the quantum computing ecosystem expands in scope and practicality, an increase in providers, including potentially unreliable third-party vendors, offering quantum computing as a service is anticipated. This situation presents multiple security challenges. Some emerging providers might attract customers with the promise of cheaper access to quantum computing resources and reduced waiting times. However, if the security measures of these services are not rigorously assessed, they could introduce significant risks. This scenario is similar to the risk of employing untrusted compilers for constructing quantum circuits, which could potentially expose or compromise proprietary information \cite{ saki2021split, hrda2023confidential}. Additionally, quantum-specific challenges such as crosstalk noise present significant security threats, as adversaries can exploit it to interfere with and manipulate quantum computations on cloud-based platforms \cite{harper2024crosstalk}.


Furthermore, the shared nature of cloud quantum computing resources raises concerns over intellectual property (IP) security. Quantum computers are susceptible to attacks like fault injection in multi-tenant computing environments. Compute performance can also be degraded for denial-of-service attacks if third-party calibration services provide inaccurate error rates of qubits or if the qubits are miscalibrated. Access to trustworthy quantum computing providers often involves long wait times and high costs, tempting users to opt for more affordable, readily available alternatives that may compromise security, risking intellectual property theft or tampering with computation results. The rise of efficient yet unreliable compilation services poses a significant threat to the integrity of quantum circuits by potentially introducing malicious code \cite{upadhyay2022robust, saki_survey_2021}. Despite quantum computing's capacity to address strategically critical problems involving sensitive data, its security and privacy aspects remain underexplored.

\subsubsection{Recent advances in securing and privacy-preserving quantum cloud computing}

Advancements in techniques for securing and ensuring data privacy for quantum cloud computing draw from quantum-specific and traditional cloud security measures, focusing on robust authentication, visibility, scalability, and efficient certificate management. The challenge lies in establishing secure, multi-tenant quantum cloud infrastructures that balance scalability with privacy and security considerations \cite{Nejabati:22}. Innovations in quantum mechanics have led to the proposal of various security mechanisms, including quantum key distribution \cite{BENNETT20147, PhysRevLett.67.661}, secret sharing \cite{PhysRevA.59.1829, dou2019rational}, key agreement \cite{10.1007/s10773-017-3484-6, 10.1007/s10773-017-3553-x}, direct communication \cite{WenJie2009, zhong2018analysis}, steganography \cite{qu2018novel}, teleportation \cite{tan2018perfect, liu2015improved}, sealed-bid auction \cite{liu2014attacks, liu2016multiparty}, and quantum machine learning \cite{liu2018quantum, liu2019unitary}, as well as privacy-preserving strategies for quantum databases \cite{liu2019quantum}. Table \ref{tab:qcc-security} summarizes several representative works on securing and privacy-preserving quantum cloud computing.

\begin{table}[htbp]
\caption{Representative works in securing and privacy-preserving quantum cloud computing. }
\label{tab:qcc-security}
\begin{tabular}{|l|l|l|}
\hline
\textbf{Category} & \textbf{Reference} & \textbf{Highlight} \\ \hline
\multirow{5}{*}{\begin{tabular}[c]{@{}l@{}}Securing QCC \\ Data\end{tabular}} & Broadbent et al. \cite{broadbent2009universal} & \begin{tabular}[c]{@{}l@{}}Introduced a protocol for secure cloud-based quantum computation\\ with classical clients.\end{tabular} \\ \cline{2-3} 
 & Fitzsimons et al. \cite{fitzsimons2017private, fitzsimons2017unconditionally} & \begin{tabular}[c]{@{}l@{}}Demonstrated private quantum computation using blind quantum \\ computing techniques.\end{tabular} \\ \cline{2-3} 
 & Huang et al. \cite{huang2017experimental} & \begin{tabular}[c]{@{}l@{}}Suggested a distributed approach to secure quantum circuits in \\ cloud computing environments\end{tabular} \\ \cline{2-3} 
 & Li et al. \cite{li2020verifiable} & \begin{tabular}[c]{@{}l@{}}Developed a scheme enhancing privacy and verifiability in quantum\\  cloud computing based on blind computation\end{tabular} \\ \cline{2-3} 
 & Mahadev \cite{mahadev2020classical} & \begin{tabular}[c]{@{}l@{}}Developed a homomorphic encryption scheme for quantum circuits, \\ ensuring secure delegation of quantum computations to servers.\end{tabular} \\ \hline
\multirow{4}{*}{\begin{tabular}[c]{@{}l@{}}Securing QCC \\ Platforms\end{tabular}} & Acharya et al. \cite{acharya2020lightweight} & \begin{tabular}[c]{@{}l@{}}Proposed a technique to detect malicious changes in NISQ computer \\ error rates in cloud environments.\end{tabular} \\ \cline{2-3} 
 & Ma et al. \cite{ma2022qenclave} & \begin{tabular}[c]{@{}l@{}}Introduced QEnclave for secure quantum cloud computing, enabling \\ privacy-preserving remote quantum operations via classical controls.\end{tabular} \\ \cline{2-3} 
 & Upadhyay et al. \cite{upadhyay2023trustworthy} & \begin{tabular}[c]{@{}l@{}}Proposed a heuristics method for trustworthy computing on untrusted\\  quantum cloud hardware\end{tabular} \\ \cline{2-3} 
 & Wang et al. \cite{wang2023qumos} & \begin{tabular}[c]{@{}l@{}}Proposed the QuMoS framework to safeguard QML models against \\ model-stealing attacks\end{tabular} \\ \hline
\multirow{5}{*}{\begin{tabular}[c]{@{}l@{}}Authentication \\ and Verification\end{tabular}} & Phalak et al. \cite{phalak2021quantum} & Proposed using QuPUFs for authenticating quantum cloud hardware. \\ \cline{2-3} 
 & Tannu et al. \cite{tannu2019ensemble} & \begin{tabular}[c]{@{}l@{}}Proposed the Ensemble of Diverse Mappings (EDM) method to \\ counteract the vulnerability of NISQ computers to correlated errors.\end{tabular} \\ \cline{2-3} 
 & Chen et al. \cite{chen2021experimental} & \begin{tabular}[c]{@{}l@{}}Presented an empirical study on cryptographic verification on IBM \\ quantum cloud computers.\end{tabular} \\ \cline{2-3} 
 & Yung and Chen \cite{yung2022anti} & \begin{tabular}[c]{@{}l@{}}Proposed a cryptographic verification protocol for anti-forging \\ quantum data.\end{tabular} \\ \cline{2-3} 
 & Kahanamoku-Meyer \cite{kahanamoku_meyer2023forging} & \begin{tabular}[c]{@{}l@{}}Demonstrated classical algorithms can forge quantum data, \\ challenging the protocols' integrity and verification measures.\end{tabular} \\ \hline
\end{tabular}
\end{table}

Adapting classical security technologies like homomorphic encryption \cite{gilad2016cryptonets} and secure multiparty computation \cite{peng2023rrnet, riazi2018chameleon} for quantum resilience is crucial, as conventional encryption might falter against quantum capabilities.
Research into blind quantum computing has aimed at securing quantum circuits, crucial for cloud-based scenarios where direct access to quantum systems is unavailable \cite{broadbent2009universal, fitzsimons2017private}. Broadbent et al. \cite{broadbent2009universal} proposed a universal blind quantum computation protocol that enables clients to perform quantum computations on a server in a way that preserves the privacy of their data and computation, without requiring the client to have quantum computational resources, marking a significant advancement in the field of quantum computing in its early stage. Fitzsimons and Kashefi \cite{fitzsimons2017private, fitzsimons2017unconditionally} proposed an unconditionally verifiable blind quantum computation protocol, enhancing privacy and verifiability in client-server quantum computing setups, with significant efficiency improvements and fault-tolerance thresholds, enabling entangling gates between any logical qubits with minimal overhead. Distributed approaches for protecting quantum circuits through classical interfaces have been proposed, specifically for prime factorization, relying on shared entanglement. Huang et al. \cite{huang2017experimental} conducted an experiment demonstrating blind quantum computing for entirely classical clients, showing that clients without quantum devices can perform computations securely on quantum servers using entanglement, including tasks like factorization with built-in verification for server honesty and correctness, marking a crucial step towards secure cloud quantum computing. Additionally, Li et al. \cite{li2020verifiable} introduced a verifiable quantum cloud computation scheme leveraging blind computation, addressing privacy and verifiability for clients in cloud-based quantum computing with a novel use of cluster states, enhancing the scheme's suitability for cloud architectures and promoting data confidentiality. Mahadev \cite{mahadev2018classical} introduced the first leveled fully homomorphic encryption scheme for quantum circuits, enabling classical clients to delegate quantum computations to a quantum server securely. This technique ensures that an honest server can perform the computation without revealing any information to a malicious server. It is constructed from a quantum-secure classical homomorphic encryption scheme based on the problem of learning with errors, bridging a significant gap in quantum and classical encryption methodologies.

Several works investigated the trustworthiness and security of quantum cloud platforms. Acharya et al. \cite{acharya2020lightweight} addressed the reliability challenges in NISQ computers by proposing a method to detect malicious changes in error rates that could alter quantum circuit outputs. Their lightweight approach involves inserting test points into circuits to monitor error rates relative to qubit allocations, employing superposition, classical, and un-compute tests for side-channel analysis, and offering a security enhancement for NISQ computing. Besides, delegated quantum computation (DQC) services, enabling clients with limited quantum capabilities to delegate calculations to quantum servers, face privacy challenges. QEnclave \cite{ma2022qenclave} introduces a secure, classical-controlled cloud-based quantum hardware for remote quantum process execution, bringing classical secure enclave principles to quantum computing and ensuring user privacy. Secure DQC schemes relying on quantum-safe classical communication and post-quantum cryptography are essential, with protocols based on quantum-safe trapdoor functions offering novel solutions, despite challenges in server overhead and the need for secure computation protocols \cite{fitzsimons2017private, gheorghiu2019verification, aaronson2019complexity, mahadev2018classical}. Upadhyay et al. \cite{upadhyay2023trustworthy} proposed a heuristic method for counteracting adversarial tampering and detecting compromised hardware, addressed the challenge of ensuring trustworthy computing on cloud-based quantum hardware, including those offered by untrusted vendors, by modeling adversarial tampering and its impact. They propose distributing computation shots across various hardware options to improve reliability and performance, significantly enhancing computing outcomes for both pure quantum and hybrid quantum-classical workloads, and introducing an adaptive heuristic to optimize hardware use in real-time. Besides, the susceptibility of cloud-based quantum machine learning (QML) applications to model-stealing attacks led to the proposal of the QuMoS framework \cite{wang2023qumos}, distributing the QML model across multiple quantum cloud providers to enhance security, despite the possibility of introducing new challenges for model integrity \cite{shen2022distributed}. 

Furthermore, authentication and verification of quantum computation in cloud computing environments have recently gained attention. Phalak et al. \cite{phalak2021quantum} explored security concerns in cloud-based quantum computing, proposing Quantum Physically Unclonable Functions (QuPUF) to ensure trustworthiness and security. Addressing the risk of users being allocated inferior quality quantum hardware by untrustworthy third parties or due to malicious tampering, they developed QuPUF variants based on superposition and decoherence, tested on IBM quantum hardware. Their solution demonstrates the feasibility of using QuPUFs to distinguish between quantum hardware, enhancing user trust in cloud quantum computing environments. Tannu et al.  \cite{tannu2019ensemble} proposed the Ensemble of Diverse Mappings (EDM) method to counteract the vulnerability of NISQ computers to correlated errors, improving the reliability of quantum computations. By diversifying qubit allocations across multiple trials, EDM decreases the likelihood of consistent errors, thereby enhancing the accuracy of output inference. Their approach demonstrated a significant improvement in inference quality compared to existing mapping algorithms. Besides, Intellectual property (IP) concerns arise from reversible quantum circuits and the potential for piracy and reverse engineering attacks. Techniques like extra line augmentation with quantum multiple-valued decision diagrams (QMDD) \cite{miller2006qmdd} and binary-decision diagrams (BDD) \cite{vinkhuijzen2023limdd} offer protection, though vulnerabilities may still exist due to synthesis traces. The risk of IP infringement escalates with cloud services' performance promises and the power of NISQ machines \cite{limaye2019revisiting}. Additionally, several empirical studies on cryptographic verification in quantum cloud contexts highlight the fidelity differences between lab and commercial quantum processors and propose protocols for anti-forging quantum data. Chen et al. \cite{chen2021experimental} explored a cryptographic verification scheme to ensure the authenticity of quantum computing within cloud services, differentiating between actual quantum processing and classical simulation. Yung and Cheng \cite{yung2022anti} enhanced the security of quantum cloud computing with an anti-forging quantum data verification protocol, extending the Shepherd-Bremner IQP-based model \cite{shepherd2009temporally} to safeguard against attacks and ensure outputs are genuinely from quantum hardware. This revised protocol allows encoding multiple secret strings, significantly increasing resistance to classical hacking and providing a robust method for estimating correlation functions crucial for verification, marking a significant advancement in cryptographic verification of quantum computational power. Similarly, Kahanamoku-Meyer \cite{kahanamoku_meyer2023forging} exposed vulnerabilities in IQP-based quantum tests, demonstrating the potential for classical algorithms to forge quantum data and extract secret keys, challenging the security of quantum computational advantage demonstrations. These advancements underscore the ongoing need for research in quantum cloud security, addressing both technical and conceptual challenges to safeguard quantum computations and data privacy in an evolving landscape.

\section{Open Problems and Future directions}
\begin{figure}[htbp]
    \centering
    \includegraphics[scale=0.055]{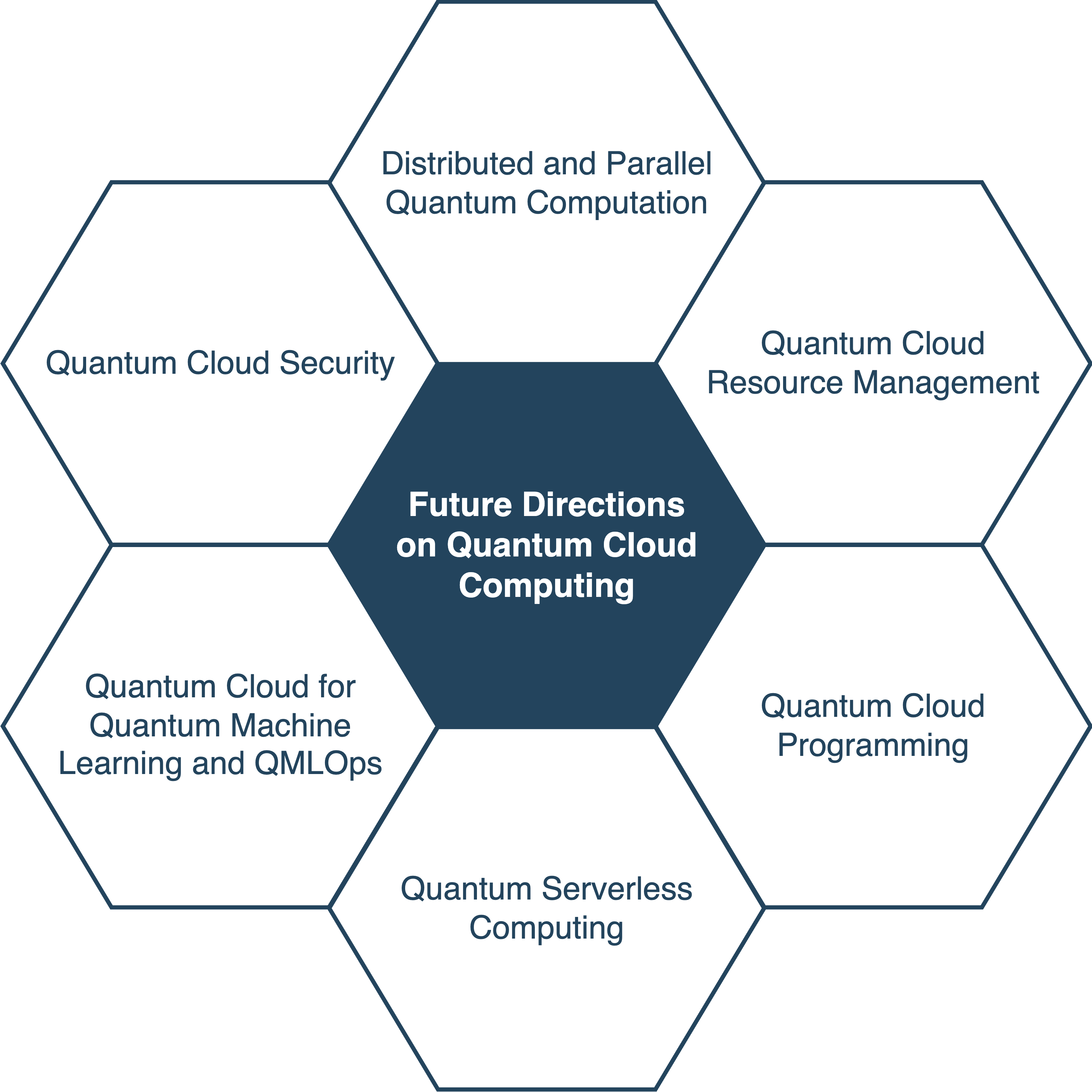}
    \caption{Open problems and future directions on Quantum Cloud Computing}
    \label{fig:qcloud-directions}
\end{figure}

Quantum cloud computing has witnessed substantial progress; however, numerous challenges remain unaddressed, necessitating further research. This section outlines critical open problems and potential future directions (as depicted in Figure \ref{fig:qcloud-directions}), aiming to advance the capabilities of quantum cloud computing.

\begin{itemize}
    \item \textbf{Distributed and Parallel Quantum Computation}: The integration of multiple quantum computing resources to harness their collective power is essential \cite{davarzani2022hierarchical, gyongyosi2021scalable}. However, this direction faces significant challenges that require extensive effort to address. 
    First, the coherence of quantum states needs to be maintained while transferring quantum data across different devices. Today's quantum computers are error-prone \cite{nisq-preskill} due to the high sensitivity of quantum states, which is easily affected by environmental factors, such as electromagnetic noise and temperature, so maintaining the qubit quality in individual quantum devices is difficult. The challenge lies in ensuring the integrity of qubits across potentially vast distances, demanding advancements in quantum networking and data communication to achieve reliable inter-device connectivity.
    Besides, the scalability of the system and the requirement of designing an effective distributed quantum algorithms are challenging for large distributed quantum systems. This requires innovative approaches to algorithm design that can navigate the intricacies of distributed quantum computations, optimizing performance across a heterogeneous network of quantum processors \cite{li2021co}. In addition, parallel quantum processing of multiple quantum tasks simultaneously on a single quantum computer is still an open problem without an efficient solution devised. Current quantum computing paradigms struggle to efficiently parallelize tasks due to the quantum decoherence and error rates associated with simultaneously manipulating multiple qubits. Developing methodologies for parallel quantum processing that can mitigate interference and maximize the utilization of quantum resources remains a critical area of research, promising to enhance the computational throughput of quantum systems significantly. Addressing these challenges is pivotal for realizing the full potential of distributed and parallel quantum computing \cite{britt2017high}.

    \item \textbf{Quantum Resource Management}: Analogous to its classical counterpart, resource management emerges as a critical issue in the quantum cloud computing landscape, attributed to the diversity of quantum hardware technologies. Due to the heterogeneity of different quantum hardware technologies, multiple aspects of the quantum resource management problem, such as allocation, scheduling, and utilization, must be addressed. It is essential to develop algorithms that can efficiently allocate and schedule resources across varied quantum tasks while optimizing their utilization \cite{nguyeniquantum2024, ravi_adaptive_2021}. 
    Quantum computation resources are inherently expensive, emphasizing the importance of designing efficient and adaptable resource allocation and scheduling algorithms. The resource utilization must also be optimized while ensuring the reliability of quantum devices to produce correct computation results. This challenge parallels those encountered in classical cloud computing, yet it is magnified by the unique complexities of quantum technology. Addressing this challenge is crucial for harnessing the full potential of quantum cloud computing, necessitating innovative approaches to navigate the intricacies of quantum resource management and facilitate the growth and application of quantum cloud computing services.
    
    \item \textbf{Quantum Cloud Programming}: The advent of quantum cloud computing has poised software engineering at the brink of significant evolution, offering access to quantum computational resources analogous to classical infrastructure. 
    However, the emerging field of quantum software development confronts several limitations, particularly when leveraging cloud-based quantum resources. A primary problem is the lack of a standardized quantum programming model. Currently, each quantum cloud provider operates on distinct software platforms, development toolkits, and standards \cite{serrano_quantum_2022}. This absence of uniformity complicates the development of cross-platform quantum applications, impeding their ability to operate seamlessly across diverse cloud environments \cite{nguyen2024qfaas}. Additionally, the deployment paradigm for quantum applications diverges significantly from that of traditional software. Unlike classical applications, which can reside on servers for on-demand invocation, quantum applications require recompilation and transmission to quantum computers for each execution. This discrepancy underscores the need for innovative approaches to quantum software lifecycle management, emphasizing the critical role of standardization and the development of deployment methodologies tailored to the quantum computing domain.

    \item \textbf{Quantum Serverless}: Leading companies in quantum computing, such as IBM, have claimed that quantum serverless is potentially the future of quantum programming \cite{ibmroadmap}. To empower the quantum serverless model in the NISQ era, some open challenges must be addressed. 
    First, resource orchestration should be optimized to deal with various heterogeneous backends and the high load of requests from multiple users. Resource orchestration refers to the collaboration of multiple processes of allocating, scheduling, and monitoring the usage of quantum and classical resources. This orchestration problem is essential for all cloud-based systems and is more challenging for serverless quantum computing systems as both quantum and classical resources are involved. They must be flexibly orchestrated to fit all the task requirements and ensure resource utilization while minimizing the total cost. 
    Second, supplement techniques, such as quantum circuit knitting \cite{piveteau2023circuit}, should be developed to enable large-scale quantum circuit execution on multiple NISQ devices. Circuit knitting can segment a large quantum circuit into smaller, manageable circuits for execution on different devices before integrating the outcomes, which is a vital area of research. Innovative approaches, including entanglement forging \cite{eddins_doubling_2022} and circuit cutting \cite{tang_cutqc_2021}, are under exploration to address these challenges. These techniques are essential in overcoming the limitations of NISQ devices, enabling the execution of complex quantum computations, and advancing the serverless quantum computing model.
    
    \item \textbf{Quantum Cloud for Quantum Machine Learning Applications and QMLOps}: In the domain of classical computing, Machine Learning Operations (MLOps) integrate machine learning workflows with DevOps practices to enhance the efficiency of ML application development \cite{john2021towards, kreuzberger2023machine}. Similarly, in the quantum computing realm, Quantum Machine Learning (QML) has emerged as a promising field, attracting significant interest from researchers \cite{houssein2022machine}. The design of cloud-based systems tailored for QML applications is essential for their advancement. Adopting an approach akin to MLOps, termed QMLOps, could significantly streamline the development and management of QML applications. By applying the principles of MLOps to the quantum context, QMLOps aims to facilitate the seamless integration, deployment, and operation of QML workflows, thereby accelerating the maturation of quantum computing technologies and their application in solving complex computational problems. This convergence of quantum computing and machine learning within cloud environments underscores the potential for QMLOps to act as a trigger for innovation and efficiency in the realm of quantum machine learning.
    
    \item \textbf{Quantum Cloud Security}: The imperative to secure data and privacy within quantum cloud computing cannot be exaggerated, especially given quantum computing's capability to compromise conventional cryptographic techniques. There is a pressing need to develop security strategies rooted in quantum principles to protect the quantum cloud ecosystem. This includes a thorough exploration and mitigation of potential quantum-specific attack vectors. Additionally, Quantum Key Distribution (QKD) \cite{mehic_quantum_2020, cao2022evolution} emerges as a pivotal research area in quantum cloud security, necessitating ongoing enhancements to ensure the secure transmission of encryption keys within cloud infrastructures. Despite progress in formulating attack models and devising corresponding defensive measures for quantum computing, current strategies only begin to address the breadth of potential security challenges. Several critical areas require further investigation, such as securing qubit technologies, quantum hardware and platforms, and preserving data security for quantum cloud applications \cite{ghosh2023primer}. Addressing these aspects of quantum cloud security is vital for advancing the field and ensuring the integrity and confidentiality of data within quantum computing environments.
    
\end{itemize}

\section{Summary and Conclusions}
In this paper, we carried out a comprehensive review of recent advances, open problems, and future directions in quantum cloud computing. We introduced emerging concepts and models of quantum cloud, such as hybrid classical-quantum cloud, quantum computing as a service, and quantum serverless. Besides, we summarized representative studies in each aspect to highlight the importance of combining quantum and cloud computing to accelerate quantum engineering. We discussed the potential applications of quantum clouds in different areas. We explored key research problems in quantum clouds, including resource management, distributed computation, and quantum cloud security. We also highlighted several research challenges and suggested future directions to advance the quantum cloud. Although there are still challenges to be addressed, quantum cloud computing has the potential to drive innovation and bring significant benefits to realize the advantage of practical quantum computing. We believe that quantum cloud computing is a promising strategy to revolutionize how quantum computing is used to solve intractable problems for classical computing in the coming years. This calls for access to quantum hardware, robust network connectivity, quantum software tools, security measures, scalability, interoperability, and cost-effectiveness.

\begin{acks}
Hoa T. Nguyen acknowledges the support from the Science and Technology Scholarship Program for Overseas Study for Master’s and Doctoral Degrees, Vin University, Vingroup, Vietnam.

\end{acks}
\bibliographystyle{ACM-Reference-Format}
\bibliography{bibliography-citedrive}
\appendix

\end{document}